  \documentclass[journal,onecolumn,draftcls,12pt]{IEEEtran}

\usepackage{cite}
\usepackage{multirow}
\usepackage{graphicx}
\usepackage{amssymb,amsmath,times,wasysym}
\usepackage{pb-diagram}
\usepackage{psfrag,balance}
\usepackage{srcltx,verbatim,color} 
\usepackage[caption=false]{subfig} 

\usepackage{xcolor}
\usepackage{mathtools}

\usepackage{algorithmic}
\usepackage{algorithm}
\usepackage[ansinew]{inputenc}

\newcounter{appendx}

\newcommand{\real}[1]{\Re\left({#1}\right)}

\renewcommand{\exp}[1]{\mathrm{e}^{#1}}

\newcommand{\Exp}[1]{\mathrm{exp}\left(#1\right)}

\newcommand{\errc}[1]{\mathrm{erfc}\left(#1\right)}

\newcommand{\diag}[1]{\mathrm{diag}\left(#1\right)}

\renewcommand{\log}[2][]{\mathrm{log}_{#1}\left(#2\right)}

\newcommand{\E}[2][]{\mathbb{E}_{#1}\!\left[{#2}\right]}

\newcommand{\Var}[1]{\mathbb{V} \mathrm{ar}\!\left[{#1}\right]}

\DeclareMathOperator{\sinc}{sinc}

\title{Performance Analysis of Distributed Intelligent Reflecting Surfaces for Wireless Communications\vspace{-1mm}}

\author{ Diluka Loku Galappaththige, \IEEEmembership{Student Member, IEEE}, Dhanushka Kudathanthirige, \IEEEmembership{Student Member, IEEE}, and  {Gayan Amarasuriya Aruma Baduge, \IEEEmembership{Senior Member, IEEE}\vspace{-5mm}}
	
\thanks{The authors are with the School of Electrical, Computer, and Biomedical Engineering, Southern Illinois University, Carbondale, IL, USA, Email: \{diluka.lg, dhanushka.kudathanthirige, gayan.baduge\}@siu.edu. This work in part has been presented at IEEE Global Communication Conference (Globecom), Dec., 2020 \cite{Diluka2020}.  }\vspace{-12mm}}
\begin{document}
\bstctlcite{IEEEexample:BSTcontrol}
\vspace{0mm}
\maketitle
 

\begin{abstract}
	In this paper, a comprehensive performance analysis of a distributed intelligent reflecting surfaces (IRSs)-aided communication system is presented. First, the   optimal signal-to-noise ratio (SNR), which   is attainable   	through the direct and reflected channels, is quantified by controlling the phase-shifts of the distributed IRSs. Next, this optimal SNR is statistically characterized by deriving tight approximations to   the  probability density function (PDF) and cumulative distribution function (CDF) for Nakagami-$m$ fading.
	Our PDF/CDF analysis is used to derive tight approximations/bounds for 
	the outage probability,   achievable rate, and average symbol error rate (SER)   in closed-form. To obtain useful insights, the asymptotic outage probability and average SER  are derived for the high SNR regime. Thereby, the achievable diversity order and  array gains are quantified. Our asymptotic performance analysis reveals that the diversity order can be boosted by using distributed passive IRSs without generating  additional electromagnetic (EM) waves via active radio frequency   chains. Our asymptotic rate analysis shows that the   lower and upper rate bounds converge to an asymptotic limit in large reflecting element regime. 
	Moreover, the transmission impairments are studied by investigating the detrimental effects  of  phase-shift quantization errors and the spatially correlated fading at the IRSs.
	Our analysis is validated via Monte-Carlo simulations. We present a rigorous set of   numerical results to investigate the performance gains of the proposed system model.  Our analytical and numerical  results reveal that the  performance of   single-input single-output wireless systems can be   boosted by recycling the EM waves   generated by a transmitter through distributed passive IRS reflections to enable constructive signal combining at a receiver.

\end{abstract}

 \linespread{1.2}

\vspace{-0mm}
\section{Introduction}\label{sec:introduction}
 \vspace{-1mm}

Recently, a novel concept of coating physical objects such as building walls and windows within a wireless propagation medium with intelligent reflecting surfaces (IRSs), which can interact with electromagnetic (EM) waves, has emerged \cite{Diluka2020,Liaskos2018,Renzo2019}.  
By invoking   the recent research developments of   meta-surfaces \cite{Sekitani2009}, the feasibility of synthesizing  the IRS has been explored   \cite{Liaskos2018}.
An IRS consists of a large number of passive reflecting elements, which can introduce controllable phase-shifts to incident EM waves. 
These phase-shifts can be intelligently controlled to enable a constructive addition of EM waves at a desired destination and thereby boosting the signal-to-noise ratio (SNR) of the  end-to-end communication.
Moreover, the IRS enables recycling of EM waves without generating additional signals via  radio-frequency (RF) chains/amplifiers and  thus enhancing the energy efficiency. 
The current state-of-the-art wireless systems, including conventional communication systems such as massive multiple-input multiple-output (MIMO), millimeter wave (mmWave) communication, and  ultra-dense deployments of small cells consume a high amount of energy and a considerable hardware implementation cost \cite{Zhang2017}. The emerging  mmWave communications may suffer from high path-losses,  penetration losses, and signal blockages. To this end, IRS-aided communication systems  can potentially address the need of  greener, more cost effective,  and more sustainable wireless connectivity  by enabling  control over the propagation environment by recycling  EM signals.
The recent developments in physics of meta-surfaces \cite{Sekitani2009} may contribute to cost effective implementation of IRS-aided wireless systems. Thus, by intelligently controlling the phase-shifts introduced by the reflecting elements, the IRS-aided system can potentially overcome unfavorable propagation conditions, increase the coverage area while retaining a higher energy efficiency  due to the passive reflecting elements. 

 \vspace{-2mm}
 \subsection{Related Prior Research on Intelligent Reflecting Surfaces}
  \vspace{-1mm}


In \cite{Liaskos2018,Sekitani2009,Basar2019,Tang2019,Renzo2019}, the initial designs and  theoretical modeling aspects of IRS-aided wireless communication are investigated. To this end, reference \cite{Liaskos2018} explores the 
core architectural designs and features of    IRS-based   wireless communications.  In \cite{Renzo2019}, a new communication-theoretic model for the analysis and optimization of IRS-aided  smart radio environments is proposed.
In \cite{Sekitani2009},  prototypes of  meta-surfaces and meta-tiles that can be used to  coat objects embedded within a smart wireless propagation environment    are already  developed. In  \cite{Basar2019}, the ray-tracing techniques are used to design novel propagation/path-loss models for IRS-aided wireless communications.  
In \cite{Tang2019}, the  path-loss models for IRS channels are designed based on the EM  properties, and, these models are categorized into far-field, near-field beamforming, and near-field broadcasting.

References \cite{Chen2019,Kudathanthirige2020,Basar2019_E,Han2019,Jung2019b,Jung2019,Hu2020,Psomas2019} provide performance analyses for IRS-enabled networks. 
Specifically, in \cite{Kudathanthirige2020}, the fundamental performance metrics of IRS-aided  communication systems operating over Rayleigh fading   are derived. 
In \cite{Chen2019},    novel IRS-aided techniques are explored to  boost the physical layer security aspects of wireless communications. 
In \cite{Basar2019_E}, the potential of the large intelligent surface (LIS)-aided communication   in boosting the      performance is investigated.  
Reference  \cite{Han2019} explores the feasibility of exploiting   statistical channel state information (CSI) to  optimize  phase-shift designs at the IRS to maximize the achievable  spectral efficiency. 
In \cite{Jung2019b}, the asymptotically achievable uplink sum rate of an IRS-aided  system    is analyzed  by deriving the rate distributions for  Rician fading channels. 
In \cite{Jung2019}, the  impact of   practically-viable/limited  IRS control channels is investigated, and a passive beamformer design, which can achieve the asymptotic optimal received SNR   under discrete reflection phase-shifts,  is presented.   
In \cite{Jung2019}, a resource allocation algorithm to maximize the  asymptotic  sum rate under the passive beamforming  is proposed. 
In \cite{Hu2020}, a performance comparison between the spherical and two-dimensional (2D)  LIS deployments   is presented, and it is shown that former can provide  wider coverage while having  simpler   positioning and flexible implementation over the 2D counterpart. 
In \cite{Psomas2019}, an analytical framework for random rotation based IRS-aided   systems is  presented, and     four low-complexity schemes via selection-based and coding-based approaches are proposed.


In \cite{Wu2019,Wu2020a,Huang2019,Abeywickrama2020,Zhang2020b,Guo2020,Cui2019,Guan2020}, the optimization-based algorithm designs are developed to intelligently control the phases of passive reflecting elements of an IRS to enable a myriad of wireless communication applications.  
In \cite{Wu2019,Wu2020a}, the joint optimization of precoders at the base-station (BS) and phase-shifts at the IRS is investigated, and the proposed solutions are based on the semidefinite relaxation   and alternating optimization techniques.
Reference \cite{Huang2019} presents two computationally efficient  algorithms to maximize the energy efficiency for  the transmit power allocation based on alternating maximization adopting
gradient descent method and sequential fractional programming approach. 
In \cite{Abeywickrama2020}, the beamforming designs are investigated with a practical phase-shift at the IRS. 
In \cite{Zhang2020b}, the capacity regions of the   IRS-aided two user multiple-access channel are characterized. Thereby, the algorithms to find the inner and outer bounds of the capacity region are   proposed for a centralized IRS deployment.  
Reference \cite{Guo2020} investigates an IRS-aided wireless  system operating in Rician fading channels by adopting  maximal ratio transmission (MRT) at the BS. 
In \cite{Cui2019}, the joint optimization of active
transmit and passive reflect beamforming is investigated to maximize the secrecy  rate for an IRS-aided secure communication system. In \cite{Guan2020}, the impact of artificial noise on IRS-aided secure communication  is investigated.

	In \cite{Yuwei2021}, a low-complexity passive reflective beamforming design is proposed for a multiple-input single-output system having distributed IRSs. By exploiting the statistical CSI, the achievable rate of the proposed system is analyzed. 
	Reference \cite{Mei2021} studies IRS selection and beam-routing for multiple IRSs-aided MIMO systems by maximizing the received signal powers at users through active and passive beamforming designs at the BS and IRSs, respectively.
	In \cite{Mei2021a}, the concept of IRS selection and beam-routing in \cite{Mei2021} is extended by adopting practical codebooks for optimizing active and passive beamforming. 
	Authors in \cite{Liang2020} propose two statistical distributions to approximate the channel distributions of two different dual-hop IRS-based wireless systems, namely, dual-hop IRS-aided scheme and IRS-aided transmit scheme. 
	In \cite{Liang2021},  the asymptotic outage probability and the asymptotic sum rate are analyzed for a multiple IRSs-aided system over Rayleigh fading.

This paper goes well beyond our related conference paper   \cite{Diluka2020} by presenting the average symbol error rate (SER), asymptotic outage/SER analysis in high SNR regime,  diversity order, array/coding gain, achievable rate limits in  large reflecting element regime, transmit power scaling laws,  effects of transmission impairments of  phase-shift quantization errors and  correlated fading for the distributed IRSs-aided communication systems.

 \vspace{-2mm}
 \subsection{Motivation and Our Contribution}
 \vspace{-1mm}
 In this subsection, we explicitly highlight the motivation and distinct contribution of this work. 

\begin{enumerate}
	\item    The main contribution of this work is to derive performance metrics for the distributed IRSs deployment over Nakagami-$m$ fading in closed-form. The detrimental impacts of practical transmission impairments, including the phase-shift quantization errors and correlated fading, are investigated, and thereby, useful design insights are obtained through numerical results. 
	
	\item A common attribute of the above related prior works \cite{Basar2019,Wu2019,Wu2020a,Kudathanthirige2020,Han2019,Chen2019,Tang2019,Basar2019_E,Zhang2020b,Jung2019,Jung2019b,Guo2020,Hu2020,Psomas2019,Huang2019,Abeywickrama2020} is that their system models consist of a single IRS. However, the concept of IRS was originally envisioned to coat physical objects  distributed within a wireless propagation channel. The fundamental performance metrics of a distributed IRSs-aided   set-up have not yet been explored.
	
	\item   To fill this important gap in IRS literature,   we present a statistical characterization of the end-to-end optimal SNR   and the corresponding performance metrics of a distributed IRSs-aided system   over Nakagami-$m$ fading. The optimal SNR is tightly approximated by using a mathematically tractable alternative through the central limit theorem (CLT) \cite{papoulis02} because the optimal counterpart does not seem to be amenable to closed-form characterization of the performance metrics.  
	Then, the  probability density function (PDF) and cumulative distribution function (CDF) of this   optimal SNR approximation are derived.

	\item By using our PDF/CDF analysis, the outage probability,  average SER, and   achievable rate bounds/approximations are derived in closed-form. In order to obtain further insights, we resort to an asymptotic analysis of the outage probability and average SER in the high SNR regime through the first order polynomial expansions of the PDF of the corresponding random variables (RVs). Since we have not used the CLT approximation for our asymptotic analysis, our high SNR metrics are asymptotically accurate, and hence, they are used to quantify the  
	achievable diversity order and array/coding gain in closed-form.

	\item Our asymptotic analysis reveals that the achievable diversity order can be linearly increased with the product of the   number of distributed  IRSs and the number of  elements in each IRS.
	Thus,  diversity gains can also be  achieved via passive reflections at the   distributed IRSs deployment by recycling existing EM waves without generating additional EM waves by active RF chains.  Our  rate analysis in the large reflecting  element regime reveals that our lower and upper bounds for the achievable rate are asymptotically accurate.

	\item The impact of  erroneously estimated phase-shifts is also quantified by adopting quantized phase-shifts at the IRSs. 
	Moreover, the  detrimental effect of spatial correlation at the reflecting elements  is explored for the proposed distributed IRS set-up.  
	
	\item The accuracy of our analysis is verified via Monte-Carlo simulations, and a set of rigorous numerical results is presented to investigate the performance of the distributed IRSs. Our results reveal that a distributed deployment of IRSs can be exploited to boost the outage,  average SER, and achievable rate performances compared to the direct transmission and single IRS-aided set-ups. We reveal that the practical transmission impairments such as  phase-shift quantization errors and spatially correlated fading can hinder the performance gains. To this end, our analysis can be useful in quantifying the amount of performance degradation due to these practical impairments. Moreover, the performance gains of the distributed IRS set-up as compared to a single/large IRS with the same number of reflective elements depend on the locations of  the IRSs. 
\end{enumerate}


\noindent
\textbf{Notation:} $\mathbf x^{T}$ denotes the transpose of  $\mathbf x$.
$\E[]{X}$ and $\Var{X}$ represent the expectation and variance of  a RV $X$, respectively.
$X\sim\mathcal {N}\left(\mu_X,  \sigma_X^{2}  \right) $ denotes that $X$ is Gaussian distributed with $\mu_X$ mean and $\sigma_X^{2}$ variance.
$\mathcal O(\cdot)$ denotes the higher order terms of the  MacLaurin series expansion \cite[Eq. (0.318.2)]{Gradshteyn2007}. $\mathcal{Q}(\cdot)$ is the Gaussian-$\mathcal{Q}$ function \cite{papoulis02}, $\Gamma(t)$ is the Gamma function \cite[Eq. (8.310.1)]{Gradshteyn2007}, $\Gamma(\alpha,x) $ is the upper incomplete Gamma function \cite[Eq. (8.350.2)]{Gradshteyn2007}, and  $\gamma(\alpha,x) $ denotes the lower incomplete Gamma function \cite[Eq. (8.350.1)]{Gradshteyn2007}. Moreover,  $\mathcal{K}_{v}(\cdot)$ represents the modified Bessel function of the second kind \cite[Eq.  (8.407)]{Gradshteyn2007} and  $F(\cdot,\cdot;\cdot;\cdot)$ is the Gauss Hypergeometric function \cite[Eq. (9.100)]{Gradshteyn2007}. Furthermore, $\mod{\!(\cdot, \cdot)}$ denotes the modulo operator and $\lfloor \cdot \rfloor$ truncates the argument \cite{Emil2020a}

\begin{figure}[!t]\centering \vspace{-7mm}
	\def\svgwidth{230pt} 
		\fontsize{8}{4}\selectfont 
	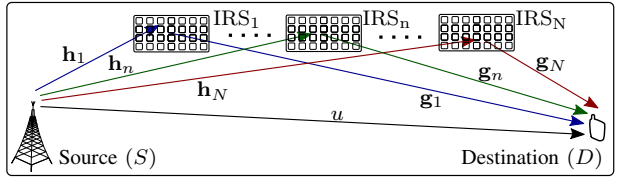 \vspace{-4mm}
	\caption{System model - a distributed IRSs-aided communication set-up. }\label{fig:system_model} \vspace{-10mm} 
\end{figure}

\vspace{-0mm}
\section{System, Channel and Signal Models  }\label{sec:system_model}

\subsection{System and Channel Model}\label{sec:system_and_channel}

We consider a distributed IRSs-aided wireless system   having a single-antenna source ($S$), a single-antenna destination ($D$), and $N$ distributed IRSs with $L$ passive reflective elements at each IRS (see Fig. \ref{fig:system_model}). 
 We assumed that   $S$ and $D$ are placed in the far-field of the IRS.
The $S$ serves  $D$ through $N$ distributed IRSs and via direct channel. 
 It is assumed that the phase-shifts  at the IRS reflective elements can be intelligently controlled such that the received signal at $D$ can be constructively added by using an IRS controller, which connects the IRSs and the BS via a backhaul connection \cite{Wu2019,Wu2020a}.  
The  channels between $S$ and the $l$th reflective element in the $n$th IRS and  the $l$th reflective element in the $n$th IRS and $D$ are denoted by $h_{nl}$ and $g_{nl}$, respectively. Moreover, $u$ denotes the direct channel between $S$ and $D$.  
All channel envelopes are assumed to be independent Nakagami-$m$ distributed, where $m$ is the shape parameter. The direct channel ($u$) can be written in its polar form as
\vspace{-2mm} 
\begin{eqnarray}\label{eqn:dirct_chnl}
	u = \alpha_u \exp{j\theta_{u}},
\end{eqnarray}

\vspace{-3mm}
\noindent where $\alpha_u$ is the envelope of $u$  and $\theta_{u}$ is the phase of $u$. The PDF of $\alpha_u$ is given by \cite{papoulis02}
\vspace{-0mm}
\begin{eqnarray}\label{eqn:dirct_chnl_pdf}
	f_{\alpha_u}(x) = \frac{2m_u^{m_u} x^{2m_u-1}}{\Gamma(m_u)\xi_u^{m_u}} \Exp{\frac{-m_u x^2}{\xi_u}},
\end{eqnarray}

\vspace{-1mm}
\noindent where $m_u$ is the shape parameter and $\xi_u\!=\!m_u\zeta_u$ is the scaling parameter. Here, $\zeta_u$ accounts for the large-scale fading/path-loss.  The $S$-to-IRS and IRS-to-$D$   channels can be defined as
\vspace{-1mm} 
\begin{eqnarray}\label{eqn:channels}
	v_{nl} = \alpha_{v_{nl}} \exp{j\theta_{v_{nl}}},
\end{eqnarray}

\vspace{-3mm}
\noindent where $v \!\in\! \{h,g\}$, $\alpha_{v_{nl}}$ is the envelope of $v_{nl}$, and $\theta_{v_{nl}}$ is the phase. The PDF of $\alpha_{v_{nl}}$ is given as 
\vspace{-0mm}
\begin{eqnarray}\label{eqn:chnl_pdf}
	f_{\alpha_{v_{nl}}}(x) = \frac{2m_v^{m_v} x^{2m_v-1}}{\Gamma(m_v) \xi_{v_n}^{m_v}} \Exp{\frac{-m_v x^2}{\xi_{v_n}}},
\end{eqnarray}

\vspace{-1mm}
\noindent where $m_{v_{nl}}=m_v$ and $\xi_{v_{nl}}=m_{v_{nl}}\zeta_{v_{nl}} = \xi_{v_{n}}$ are the shape and scaling parameters, respectively. It is assumed that large-scale fading parameters  are the same for a given IRS because its reflective elements are co-located; $\xi_{v_{nl}}=\xi_{v_{n}},\; \forall l$. However,  $\xi_{v_{n}}$ depends on the IRS index ($n$) because   geographically distributed IRSs deployment.

\subsection{Signal Model}\label{sec:sgnl_modl}

The signal transmitted by $S$ reaches at $D$ through the reflected channels of $N$ distributed IRSs and the direct channel. The signal received at $D$ can be written as
\vspace{-1mm}  
\begin{eqnarray}\label{eqn:rx_signl}
	r =\sqrt{P}\left(u + \sum\nolimits_{n=1}^{N} \mathbf{g}^T_n \mathbf \Theta_n \mathbf{h}_n \right) x + w,
\end{eqnarray}

\vspace{-1mm}
\noindent where $x$ is the transmitted signal at $S$ satisfying $\E{|x|^2} =1$,  $P$ is the transmit power, and $w$ is an additive white Gaussian noise (AWGN) at $D$ having zero mean and variance of  $\sigma_{w}^2$ such that  $w\sim \mathcal{CN}(0,\sigma_{w}^2)$. In \eqref{eqn:rx_signl}, $\mathbf{h}_n = [h_{n1},\cdots, h_{nl}, \cdots, h_{nL}]^T\in \mathbb C^{L\times 1}$ is the channel vector between $S$ and the $n$th IRS  and  $\mathbf{g}^T_n = [g_{n1},\cdots, g_{nl}, \cdots, g_{nL}]\in \mathbb C^{1\times L}$ accounts for  the channel vector between  the $n$th IRS and $D$. Moreover, $\mathbf \Theta_n= \diag{\eta_{n1} \exp{j\theta_{n1}}, \cdots, \eta_{nl} \exp{j\theta_{nl}}, \cdots, \eta_{nL} \exp{j\theta_{nL}}}\in \mathbb C^{L\times L}$ is a diagonal matrix, which  captures the reflection properties of the $n$th IRS. Here, $\eta_{nl} \exp{j\theta_{nl}}$ is a complex-valued reflection coefficient at the $l$th reflective element of the $n$th IRS, where $\eta_{nl}$ and $\theta_{nl}$  are the magnitude of attenuation and phase-shift, respectively.
Thus, the  signal received  at $D$ in  \eqref{eqn:rx_signl} can be rewritten as
\vspace{-1mm}
\begin{eqnarray}\label{eqn:rx_signl_rearng}
	r = \sqrt{P}\left(u + \sum\nolimits_{n=1}^{N} \sum\nolimits_{l=1}^{L} g_{nl} \eta_{nl} \exp{j\theta_{nl}} h_{nl} \right) x + w.
\end{eqnarray}

\vspace{-1mm}
\noindent The SNR at $D$  can be derived via \eqref{eqn:rx_signl_rearng} as
\vspace{-1mm}
\begin{eqnarray}\label{eqn:snr}
	\gamma = \bar{\gamma} \left|u + \sum\nolimits_{n=1}^{N} \sum\nolimits_{l=1}^{L} g_{nl} \eta_{nl} \exp{j\theta_{nl}} h_{nl} \right|^2,
\end{eqnarray}

\vspace{-1mm}
\noindent where $\bar{\gamma}=P/\sigma^2_w$ is the transmit SNR. 
Then, by substituting \eqref{eqn:dirct_chnl} and  \eqref{eqn:channels} into \eqref{eqn:snr}, this SNR can be written in terms of the  channel phases as
\vspace{-1mm}
\begin{eqnarray}\label{eqn:snr_rew}
	\gamma = \bar{\gamma} \left|\alpha_u \exp{j \theta_{u}} + \sum\nolimits_{n=1}^{N} \sum\nolimits_{l=1}^{L} \eta_{nl} \alpha_{g_{nl}} \alpha_{h_{nl}}  \exp{j\left(\theta_{nl} + \theta_{g_{nl}} + \theta_{h_{nl}}\right)}  \right|^2.
\end{eqnarray} 	

\vspace{-1mm}
\noindent This   analysis \eqref{eqn:snr_rew} reveals that the received SNR at $D$ can be maximized   when $NL$  signal terms  inside the summation of \eqref{eqn:snr_rew} are constructively added to the signal component received via the direct channel. By controlling the phase-shifts at each IRS reflective element ($\theta_{nl}$), the phases inside the double summation in \eqref{eqn:snr_rew} can be  adjusted    to enable a constructive addition of the received signal components via the  direct   and reflected channels. 
Thus, the optimal choice of $\theta_{nl}$ to maximize the received SNR at $D$ can be given by \cite{Wu2019}
\vspace{-3mm} 
\begin{eqnarray}\label{eqn:opt_theta}
	\theta_{nl}^* =\underset{-\pi\leq \theta_{nl} \leq \pi}{ \mathrm{argmax}} \;{\gamma} = \theta_u - \left(\theta_{g_{nl}} + \theta_{h_{nl}}\right),
\end{eqnarray} 
for $n \in \{1,\cdots,N\}$ and $l \in \{1,\cdots,L\}$. By using \eqref{eqn:opt_theta}, the optimal SNR at $D$ can be derived as
\vspace{-1mm}
\begin{eqnarray}\label{eqn:snr_opt}
	\gamma^*= \bar{\gamma} \left[\alpha_u  + \sum\nolimits_{n=1}^{N} \sum\nolimits_{l=1}^{L} \eta_{nl} \alpha_{g_{nl}} \alpha_{h_{nl}}   \right]^2.
\end{eqnarray}

\section{Preliminary Analysis}\label{sec:Preliminary_analysis}
 
\subsection{Statistical Characterization of the Optimal Received SNR }\label{sec:stat_chr_gamma}

The $\alpha_{g_{nl}}$ and $\alpha_{h_{nl}}$ in (\ref{eqn:snr_opt}) for $n \in \{1,\cdots,N\}$ and $l \in \{1,\cdots,L\}$ are independently distributed Nakagami RVs, and hence,  the exact derivations of the PDFs of 
\vspace{-1mm}
\begin{eqnarray}\label{eqn:R_approx}
	Y = \sum\nolimits_{n=1}^{N} \sum\nolimits_{l=1}^{L} \eta_{nl} \alpha_{g_{nl}} \alpha_{h_{nl}}, \quad \text{and} \quad  {\gamma}^* = \bar{\gamma} { [\underbrace{\alpha_u + Y  }_{R} ]}^2
=\bar{\gamma}R^2,
\end{eqnarray}

\vspace{-1mm}
\noindent seem  mathematically involved and may not provide useful design insights. Nevertheless, even for moderately large values of the product $NL$,   $Y$ can be tightly approximated by an one-sided Gaussian distributed RV ($\tilde Y$)   by invoking the CLT  \cite{papoulis02}. Then, an approximated PDF for  $Y$ can be written as    (see Appendix \ref{app:Appendix0})
\vspace{-1mm}
\begin{eqnarray}\label{eqn:pdf_Y}
f_Y(y) \approx 	f_{\tilde Y}(y) =
\frac{\psi}{\sqrt{2 \pi \sigma_{Y}^2}} \Exp{-\frac{(y-\mu_Y)^2}{2 \sigma_{Y}^2}}, \quad  y \geq 0,  
\end{eqnarray}

\vspace{-1mm}
\noindent where $f_{\tilde Y}(y)=0$ for $y<0$, $\psi \triangleq 1/\mathcal{Q}\left(-\mu_Y/\sigma_{Y}\right)$ is a normalization factor to satisfy  $\int_{-\infty}^{\infty} f_{\tilde Y}(x) dx=1$.
In (\ref{eqn:pdf_Y}),    $\mu_Y$ and   $\sigma_{Y}^2$  can be derived via the moment  matching technique    as   (see Appendix \ref{app:Appendix0})
\vspace{-1mm}
\begin{subequations}
\begin{eqnarray} \label{eqn:mean_&_var}
	\mu_Y &=& \sum\nolimits_{n=1}^{N} \sum\nolimits_{l=1}^{L}  \eta_{nl} \sqrt{\frac{\xi_{h_n} \xi_{g_n}}{m_h m_g}} \frac{\Gamma\left(m_h + 1/2\right) \Gamma\left(m_g+1/2\right)}{\Gamma\left(m_h\right) \Gamma\left(m_g\right)} ,\label{eqn:mean}\\
	\sigma_{Y}^2 &=& \sum\nolimits_{n=1}^{N} \sum\nolimits_{l=1}^{L} \eta_{nl}^2  \left({\frac{\xi_{h_n} \xi_{g_n}}{m_h m_g}} \right)  \frac{\Gamma\left(m_h+1\right) \Gamma\left(m_g+1\right)}{\Gamma\left(m_h\right) \Gamma\left(m_g\right)} -\mu_Y^2. \label{eqn:var}
\end{eqnarray}  		 
\end{subequations}

\vspace{-1mm}
\noindent 
Then, a tight approximation of the  PDF of $R$ in (\ref{eqn:R_approx}) or the  PDF of its approximation $\tilde R = \alpha_u + \tilde Y $ can be derived as (see Appendix \ref{app:Appendix1})
\vspace{-1mm}
\begin{eqnarray}\label{eqn:pdf_R}
	\!\!\!	\!\!\!\!\!\!f_R(x) \approx f_{\tilde R}(x)  = \lambda \exp{-\Delta \left(\frac{x-\mu_Y}{2\sigma_{Y}^{2} \sqrt{a}}\right)^{\!2}}  \sum_{k=0}^{2m_u\!-\!1} \!\! \binom{2m_u\!-\!1}{k} \! \! \left(\frac{x\!-\!\mu_Y}{2\sigma_{Y}^2 \sqrt{a}}\right)^{\!\!2m_u\!-\!1\!-\!k}\!\!\left[\Gamma \!\left(\!\frac{k\!+\!1}{2}, \!\left(\!\frac{x\!-\!\mu_Y}{2\sigma_{Y}^{2} \sqrt{a}}\right)^{\!\!2}\right) \right].
\end{eqnarray} 

\vspace{-1mm}
\noindent 
Here, $a$, $\lambda$,  and $\Delta$ are  given by
\vspace{-1mm}
\begin{eqnarray} \label{eqn:def_1}
	\!\!\!\!\!\! a &=& \frac{m_u}{\xi_u} + \frac{1}{2\sigma_{Y}^2},  \label{eqn:def_a} \quad \text{and} \quad \lambda = \frac{m_u^{m_u} \psi }{ \Gamma(m_u) \xi_u^{m_u} a^{m_u}\sqrt{2\pi \sigma_{Y}^2}}, \label{eqn:def_lambda} \quad \text{and} \quad 
	\Delta = \left(\frac{1}{2\sigma_{Y}^2} - \frac{1}{4a \sigma_{Y}^4 }\right) 4\sigma_{Y}^4 a. \label{eqn:def_delta}	
\end{eqnarray}  		 

\vspace{-1mm}
\noindent Via $\tilde R = \alpha_u + \tilde Y $  in (\ref{eqn:pdf_R}), a tight approximation to the optimal SNR ($\gamma^*$) can be written as 
\vspace{-2mm} 
\begin{eqnarray}\label{eqn:SNR_approx}
	\gamma^* \approx \tilde{\gamma}^* = \bar \gamma \tilde {R}^2.
\end{eqnarray}

\vspace{-2mm}
\noindent Then, an approximation for the PDF of $\gamma^* = \bar \gamma R^2 $ or the  PDF of  $\tilde{\gamma}^* =\bar \gamma \tilde {R}^2 $ in (\ref{eqn:SNR_approx})  can be derived    by using (\ref{eqn:pdf_R}) as \cite{papoulis02}
\vspace{-1mm}
\begin{eqnarray}\label{eqn:pdf_gamma}
	f_{\gamma^*}(y) &\approx& 
	f_{\tilde R}\left(\sqrt{{y}/{\bar{\gamma}}}\right) \Big/ {2 \sqrt{\bar{\gamma} y}}.
\end{eqnarray} 

\vspace{-1mm}
\noindent Next, the CDF of $\tilde R$ can be derived as (see Appendix \ref{app:Appendix2})
\vspace{-1mm}
\begin{eqnarray}\label{eqn:cdf_R}
	F_{\tilde{R}}(x) &=&  1 - \int_{x}^{\infty} f_{\tilde{R}}(u) du= 1 - \frac{\lambda }{2 a^{m_u}} \sum_{k=0}^{2m_u-1} \binom{2m_u-1}{k} I_{k},
\end{eqnarray} 

\vspace{-1mm}
\noindent where $I_{k}$ is given as
\vspace{-1mm}
\begin{eqnarray}\label{eqn:I_mu}
	I_{k}  =\begin{cases}
	I_{k}^{o},  & \quad   \text{odd $k$ },  \\
	I_{k}^{e},  &  \quad   \text{even $k$} .
\end{cases}
\end{eqnarray} 

\vspace{-1mm}
\noindent In \eqref{eqn:I_mu}, $I_{k}^o$ is given by
\vspace{-1mm}
\begin{eqnarray}\label{eqn:I_mu_odd}
	\!\!\!\!\! I_{k}^o \!=\!\! \begin{cases}
	\!\frac{q \left(\gamma_o\!-\!1\right)!}{2} \!\!\sum\limits_{i=0}^{\gamma_o\!-\!1} \!\! \frac{\left(\Delta+1\right)^{\frac{k}{2}-m_u-i}}{i! } \! \left(2\Gamma\!\left(m_u\!+\!i+\!\frac{k}{2}\right) - \Gamma\!\left(m_u\!+\!i\!-\!\frac{k}{2}, \left(\Delta\!+\!1\right)l_{min}^2\right) \right),  &\text{for} \quad x<\mu_{Y}, \\
	\!\frac{q \left(\gamma_o-1\right)!}{2} \sum\limits_{i=0}^{\gamma_o-1} \frac{\left(\Delta+1\right)^{\frac{k}{2}-m_u-i}}{i!}  \Gamma\left(m_u+i-\frac{k}{2}, \left(\Delta+1\right)l_{min}^2\right),   &\text{for} \quad x>\mu_{Y},
	\end{cases}
\end{eqnarray}

\vspace{-1mm}
\noindent where $q = 2\sigma_{Y}^2 \sqrt{a}$, $\gamma_o = (k+1)/2$, and $l_{min} = (x-\mu_{Y})/q$. Here, $I_{k}^e$ in  \eqref{eqn:I_mu} can be given as
\vspace{-1mm}
\begin{eqnarray}\label{eqn:I_mu_even}
	\!\!\!\!\!\!\!\!\!	I_{k}^e &=& \frac{q(\gamma_e-1)!}{2} \sum\limits_{j=0}^{\gamma_e-1} \frac{\Delta^{j-\gamma_e}}{j!}  \left( \Gamma\left(\frac{k+1}{2},l_{min}^2\right) l_{min}^{2j} \exp{-\Delta l_{min}^2}  - \frac{\Gamma\left(j+\frac{k}{2}+\frac{1}{2}, (\Delta+1)l_{min}^2\right)}{(\Delta+1)^{j+\frac{k}{2}+\frac{1}{2}}} \right), 
\end{eqnarray}

\vspace{-1mm}
\noindent where $\gamma_e = m-k/2$. Then, the CDF of $\gamma^* = \bar \gamma R^2$ can be approximated as \cite{papoulis02}
\vspace{-1mm}
\begin{eqnarray}\label{eqn:cdf_SNR}
	F_{\gamma^*}(y) &=&  \mathrm{Pr}\left(\gamma^*\leq y\right) \approx F_{\tilde R}\left(\sqrt{y/\bar{\gamma}} \right).
\end{eqnarray}

\vspace{-1mm}


\begin{figure}[!t]\centering\vspace{-7mm}
	\includegraphics[width=0.45\textwidth]{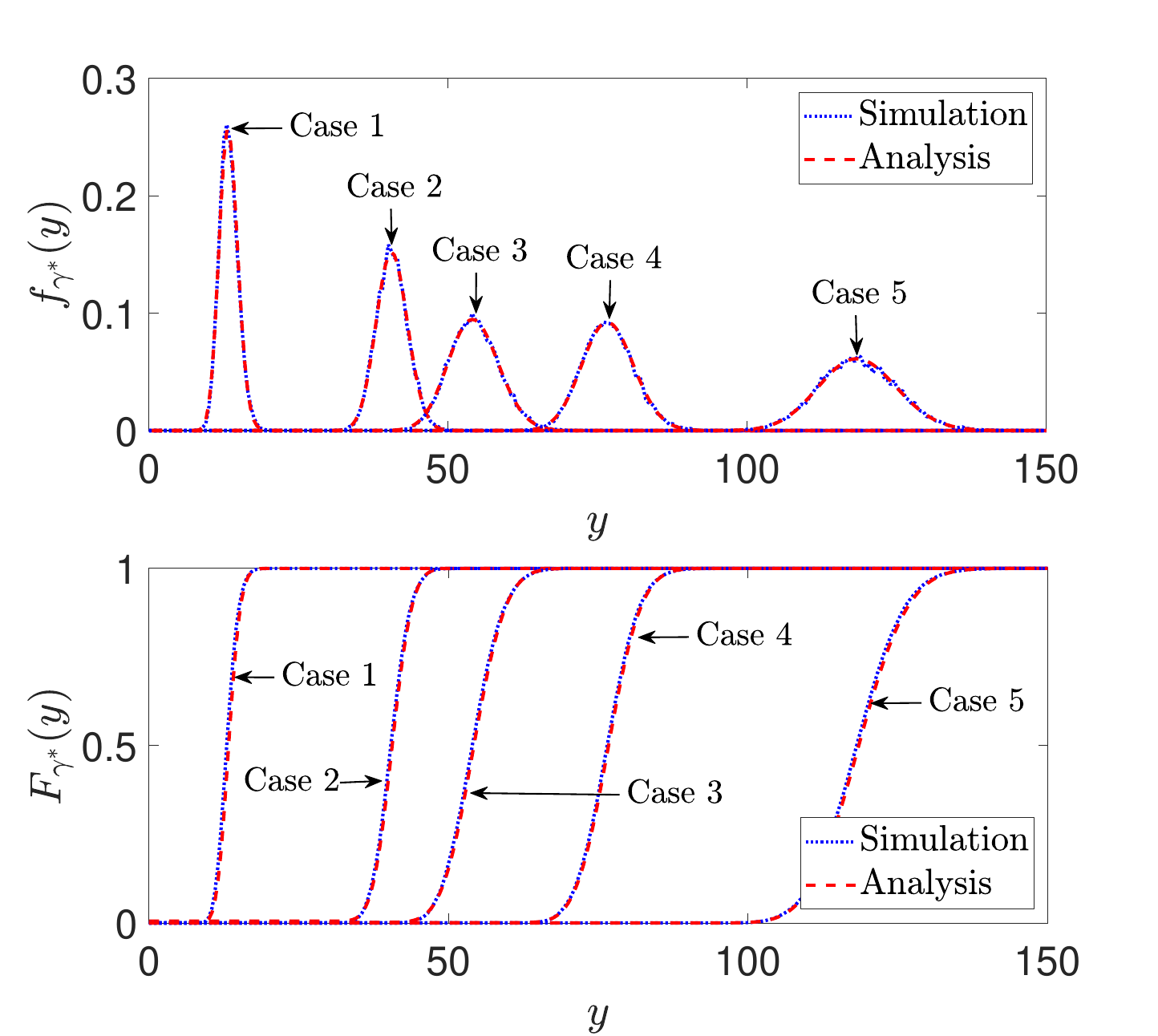}\vspace{-4mm}
	\caption{The PDF and CDF of the received SNR ($\gamma^*$) for $\bar{\gamma}= -10$dB. The combinations of $N$ and $L$ for Case 1 to Case 5 are set to $\{N=2,L=32\}$, $\{N=6,L=32\}$, $\{N=2,L=64\}$, $\{N=8,L=32\}$, and $\{N=2,L=128\}$, respectively. }
	\label{fig:pdf_cdf_N}\vspace{-5mm}
\end{figure}


\noindent  \textbf{\textit{Remark 1:}}  
The accuracy of the approximated PDF and CDF of $\gamma^*$  is verified by plotting  (\ref{eqn:pdf_gamma}), (\ref{eqn:cdf_SNR}) and the Monte-Carlo simulations of the exact counterparts in    Fig. \ref{fig:pdf_cdf_N}. It  reveals that  our analytical approximations for the  PDF (\ref{eqn:pdf_gamma}) and CDF (\ref{eqn:cdf_SNR}) of $\gamma^*$  are accurate even for moderately large number of reflective elements $(L)$  at the IRSs. Since IRSs are typically made of cost-effective reflective elements \cite{Wu2020}, a moderately large $N$ is practically feasible.

\section{Performance Analysis}\label{sec:Performance_analysis}

\subsection{Outage Probability }\label{sec:outage_prob}

The probability that the instantaneous SNR falls below a threshold SNR ($\gamma_{th}$) is defined as the SNR outage probability, and   a tight   approximation to it can be obtained   from (\ref{eqn:cdf_SNR}) as
\vspace{-2mm}
\begin{eqnarray}\label{eqn:out_prob}
P_{out} = P_{r}\left(\gamma \leq \gamma_{th}\right) \approx F_{\gamma^*}(\gamma_{th}).
\end{eqnarray}  

\vspace{-3mm}
\noindent This definition for $P_{out}$ is used to quantify the asymptotic outage probability and the achievable diversity order in Section \ref{sec:diversity}.

\subsection{Average Achievable Rate }\label{sec:achvble_rate}
The average achievable rate   can be defined as follows:
\vspace{-2mm}
\begin{eqnarray}\label{eqn:avg_rate}
	\mathcal{R} = \E{\log[2]{1+\gamma^*}} .
\end{eqnarray}   

\vspace{-2mm}
\noindent Since the exact derivation of  $\mathcal{R}$ in \eqref{eqn:avg_rate} seems mathematically intractable,  we resort to  tight upper and lower bounds by invoking the Jensen's inequality  as \cite{Zhang2014}
 \vspace{-2mm}
\begin{eqnarray}\label{eqn:rate_bound}
	\mathcal{R}_{lb} \leq \mathcal{R} \leq \mathcal{R}_{ub}.
\end{eqnarray} 

\vspace{-1mm}
\noindent In \eqref{eqn:rate_bound}, $\mathcal{R}_{lb}$ and $\mathcal{R}_{ub}$ are defined as
\vspace{-2mm}
\begin{subequations}
\begin{eqnarray} 
	\mathcal{R}_{lb} &=& \log[2]{1+ \left(\E{1/\gamma^*}\right)^{-1}}\approx \log[2]{1+ \left(\E{1/\tilde{\gamma}^*}\right)^{-1}}, \label{eqn:rate_lb}\\
	\mathcal{R}_{ub} &=& \log[2]{1+ \E{\gamma^*}}\approx \log[2]{1+ \E{\tilde {\gamma}^*}}. \label{eqn:rate_ub}	
\end{eqnarray}  		 
\end{subequations}

\vspace{-2mm}
\noindent The expectation term in \eqref{eqn:rate_ub} can be approximately derived as (see Appendix \ref{app:Appendix3_1})
\vspace{-1mm}
\begin{eqnarray}\label{eqn:E_gamma_ub}
\E{\gamma^*} \approx \E{\tilde{\gamma}^*} = \bar{\gamma} \left(\sigma_{u}^2 +\sigma_{Y}^2 + 2\mu_{u} \mu_{Y} + \mu_{u}^2 + \mu_{Y}^2 \right),
\end{eqnarray} 

\vspace{-1mm}
\noindent where $\mu_{Y}$ and $\sigma_{Y}^2$ are defined in \eqref{eqn:mean} and \eqref{eqn:var}, respectively. In (\ref{eqn:E_gamma_ub}), $\mu_{u}$ and $\sigma_{u}^2$ are defined as
\vspace{-7mm}
\begin{subequations}
\begin{eqnarray} \label{eqn:u_mean_&_var}
	\mu_u &=& \frac{\Gamma\left(m_u + 1/2\right)}{\Gamma\left(m_u\right)}  \left({\frac{\xi_u}{m_u}}\right)^{1/2},\label{eqn:u_mean}\\
	\sigma_{u}^2 &=&  \xi_u \left(1 - \frac{1}{m_u} \left(\frac{\Gamma\left(m_u+1/2\right)}{\Gamma\left(m_u\right)}\right)^2 \right). \label{eqn:u_var}
\end{eqnarray}  		 
\end{subequations}

\vspace{-1mm}
\noindent Next, the expectation term in \eqref{eqn:rate_lb} can be computed as  
\vspace{-1mm}
\begin{eqnarray}\label{eqn:E_gamma_lb}
\E{1/\gamma^*}\approx\E{1/\tilde{\gamma}^*} = {1}\big/{\E{\tilde\gamma^*}} + {\Var{\tilde\gamma^*}}\big/{\left(\E{\tilde\gamma^*}\right)^3},
\end{eqnarray}

\vspace{-1mm}
\noindent where $\E{\tilde{\gamma}^*}$ is defined in \eqref{eqn:E_gamma_ub}. Here, $\Var{\tilde\gamma^*}$ is derived as (see Appendix \ref{app:Appendix3_2})
\vspace{-1mm}
\begin{eqnarray}\label{eqn:Var_gamma_lb}
	\Var{\tilde\gamma^*} = \bar{\gamma}^2\E{\tilde R^4} - \left(\E{\tilde\gamma^*}\right)^2,
\end{eqnarray}

\vspace{-1mm}
\noindent where $\mathbb E{[\tilde R^4]}$ is given  by 
\vspace{-1mm}
\begin{eqnarray}
	\E{\tilde R^4} &=&  \sum\nolimits_{n=0}^{4} \binom{4}{n} \E{\alpha_u^{(4-n)}} \E{\tilde Y^n}.
\end{eqnarray}

\vspace{-1mm}
\noindent Here,  $\E{\alpha_u^n}$ for $n \in \{1,2,3,4\}$ can be  derived  as
\vspace{-1mm}
\begin{eqnarray}\label{eqn:E_alpha_u_n}
	\E{\alpha_u^n} = \frac{\Gamma\left(m_u+n/2\right)}{\Gamma\left(m_u\right)} \left(\frac{\xi_u}{m_u}\right)^{n/2}.
\end{eqnarray}

\vspace{-1mm}
\noindent Moreover, $\mathbb E{[\tilde Y^n]}$ for $n \in \{1,2,3,4\}$ is given by
\vspace{-1mm}
\begin{eqnarray}\label{eqn:E_Y_n}
	\E{\tilde Y^n} = \frac{\psi}{2\sqrt{\pi}} \sum\nolimits_{i=0}^{n}  \binom{n}{i} \left( \sqrt{2 \sigma_{Y}^2}\right)^{n-i}  \mu_{Y}^i I\left(n-i, \frac{-\mu_{Y}}{2 \sigma_{Y}^2}\right), 
\end{eqnarray}

\vspace{-1mm}
\noindent where  $I(\cdot,\cdot)$ is given as
\vspace{-1mm}
\begin{eqnarray}\label{eqn:I}
	I\left(m,t\right) = \begin{cases}
	(-1)^m \gamma\left(\frac{m+1}{2}, t^2\right) + \Gamma\left(\frac{m+1}{2}\right),  &\text{for} \,\, t \leq 0,  \\
	\Gamma\left(\frac{m+1}{2}, t^2\right),& \text{otherwise}.
	\end{cases}
\end{eqnarray} 

\vspace{-1mm}
\noindent 
Finally, $\mathcal{R}_{lb}$ and $\mathcal{R}_{ub}$ can be derived  as
\vspace{-1mm}
\begin{eqnarray} 
	\!\!\!\!\!\!\! \mathcal{R}_{lb} &=& \log[2]{\!\! 1 \!+\! \frac{\bar{\gamma} \left(\sigma_{u}^2 +\sigma_{Y}^2 + 2\mu_{u} \mu_{Y} + \mu_{u}^2 + \mu_{Y}^2 \right)^3}{\sum\limits_{n=0}^{4} \binom{4}{n} \frac{\Gamma\left(m_u+\frac{(4-n)}{2}\right)}{\Gamma\left(m_u\right)} \left(\frac{\xi_u}{m_u}\right)^{(4-n)/2}  \!\!\! \frac{\psi}{2\sqrt{\pi}} \sum\limits_{i=0}^{n}  \binom{n}{i} \left( {2 \sigma_{Y}^2}\right)^{(n-i)/2}  \mu_{Y}^i I\left(n-i, \frac{-\mu_{Y}}{2 \sigma_{Y}^2}\right) }}\!\!, \label{eqn:rate_lb_sub} \\
%
	\!\!\!\!\!\!\! \mathcal{R}_{ub} &=&   \log[2]{1+ \bar{\gamma} \left(\sigma_{u}^2 +\sigma_{Y}^2 + 2\mu_{u} \mu_{Y} + \mu_{u}^2 + \mu_{Y}^2 \right)},\label{eqn:rate_ub_sub}
\end{eqnarray}

\vspace{-1mm}
\noindent where $\mu_Y$, $\sigma_Y^2$,   $\mu_u$, and $\sigma_u^2$ are defined in \eqref{eqn:mean}, \eqref{eqn:var},   \eqref{eqn:u_mean}, and  \eqref{eqn:u_var}, respectively.

\subsubsection{Asymptotic achievable rate as the number of IRS elements grows large $(L\rightarrow \infty)$}\label{sec:asym_rate}
To obtain further insights, the asymptotic achievable rate can be derived as the number of IRS elements grows without bound  $(L\rightarrow \infty)$. In this operating regime, we observe that the transmit power can be scaled inversely proportional to the square of the number of IRS elements: $\lim_{L\rightarrow \infty} P = P_E/L^2$. By using this transmit power scaling law, the lower and upper rate bounds in \eqref{eqn:rate_lb_sub} and \eqref{eqn:rate_ub_sub}, respectively,  can be shown to  approach an asymptotic limit  as follows (see Appendix \ref{app:Appendix3_3}):
\vspace{-10mm}
\begin{eqnarray}\label{eqn:rate_bound_asym}
	\lim_{L\rightarrow \infty }\mathcal{R}_{lb} \longrightarrow  \mathcal{R}^{\infty}, \qquad \text{and} \qquad \lim_{L\rightarrow \infty }\mathcal{R}_{ub} \longrightarrow  \mathcal{R}^{\infty},
\end{eqnarray}

\vspace{-1mm}
\noindent where the asymptotic achievable rate $\mathcal{R}^{\infty}$ in large $L$ regime is given by
\vspace{-1mm}
\begin{eqnarray}\label{eqn:rate_asym}
	\lim_{L\rightarrow \infty }\mathcal{R} \longrightarrow \mathcal{R}^{\infty} = \log[2]{1+ \bar{\gamma}_{E} \left( \sum_{n=1}^{N} \eta_{n} \sqrt{\frac{\xi_{h_n} \xi_{g_n}}{m_h m_g}} \frac{\Gamma\left(m_h + 1/2\right) \Gamma\left(m_g+1/2\right)}{\Gamma\left(m_h\right) \Gamma\left(m_g\right)}  \right)^2},
\end{eqnarray}

\vspace{-1mm}
\noindent where $\bar{\gamma}_{E} = P_{E}/\sigma^2_w$ and  $\eta_{n} = \eta_{nl}$, $\forall n$.

\noindent \textbf{\textit{Remark 3:}} Our asymptotic  analysis reveals that our lower \eqref{eqn:rate_lb_sub} and upper \eqref{eqn:rate_ub_sub} rate bounds converge to a common limit, and  hence, these rate bounds are asymptotically accurate in large $L$ regime. Thus, a finite rate at $D$ is achievable  once 
the transmit power at $S$ can be scaled inversely proportional to $L^2$.

\noindent \textbf{\textit{Remark 4:}}  The  physical dimension of a reflecting  element is in the order of sub-wavelengths (i.e., $\lambda/10$ to $\lambda/5$, where $\lambda$ is the wavelength) \cite{Emil2020b}. Thus, thousands of reflecting elements can be embedded in a relatively smaller surface area.  To this end,  the transmission distance differences of the channels from the furthest and closest elements of a given IRS to the destination can be assumed to be the negligible in the far-field. Consequently, we assumed that the large-scale fading coefficients ($\eta_{nl}$ for $l \in \{1,\cdots,L\}$) are the same for a given IRS even for the asymptotic analysis in \eqref{eqn:rate_asym}.

\subsection{Average Symbol Error Rate (SER) }\label{sec:ber}

The average SER is the  expectation of  conditional error probability ($P_{e|\gamma^*}$) over the probability distribution of $\gamma^*$  \cite{Proakis2001}. For a wide-range of coherent  modulation schemes, $P_{e|\gamma^*}$ is given by \cite{Proakis2001}
\vspace{-10mm}
\begin{eqnarray}\label{eqn:ber}
	P_{e|\gamma^*} = \omega \mathcal{Q}\left(\sqrt{\vartheta \gamma^*}\right),
\end{eqnarray} 

\vspace{-2mm}
\noindent where $\omega$ and $\vartheta$   depend on  the modulation scheme  \cite{Proakis2001}.
For instance,    the pair of parameters  ($\omega,\vartheta$) for  binary phase-shift keying (BPSK) and quadrature phase-shift keying (QPSK) are given by ($\omega=1,\vartheta=2$) and ($\omega=2,\vartheta=1$), respectively \cite{Proakis2001}. 
 Then, the average SER can be defined as 
 \vspace{-2mm}
\begin{eqnarray}
	\bar{P}_{e} = \E{\omega \mathcal{Q}\left(\sqrt{\vartheta \gamma^*}\right)}.
\end{eqnarray}
 
 \vspace{-1mm}
 \noindent Then, by using $\tilde{\gamma}^*$ in (\ref{eqn:SNR_approx}),  a tight approximation for the  average SER can be derived as 
 \vspace{-1mm}
\begin{eqnarray}\label{eqn:ber_alt}
	\bar{P}_{e} \approx \tilde P_{e}= \omega \E[\alpha_u]{\E[\tilde{Y}]{\mathcal{Q}\left( \vartheta_b (\alpha_u+\tilde{Y})\right)}},
\end{eqnarray} 

\vspace{-1mm}
\noindent where $\vartheta_b \!=\! \sqrt{\vartheta \bar{\gamma}}$.   
A tight approximation to the   average SER   $\tilde {P}_{e}$ is derived as  (see Appendix \ref{app:Appendix3})
\vspace{-1mm}
\begin{eqnarray}\label{eqn:avg_ber}
	\!\!\!\!\!	\!\!\!\! \tilde {P}_{e} &\approx& \omega B \sum_{k=0}^{2m_u-1} \!\! \left(\frac{-u_1}{2v_1}\right)^{2m_u-1-k} \!\!\!\!\! \Gamma(\gamma_b) \sum_{i=0}^{\gamma_b-1} \!\! \frac{v_1^{i-\gamma_b}}{i! }  \left(\!\mathcal{Q}(\sqrt{2}r) \left(\frac{u_1}{2v_1}\right)^{2i} \!\!\!-\! \frac{s \exp{\frac{u_2^2}{4v_2} - \frac{u_1^2}{4v_1}-r}}{\sqrt{\pi}} \sum_{j=0}^{2i} q_b^{2i-j} I_b \right)\!, 
\end{eqnarray}

\vspace{-1mm}
\noindent where $\gamma_b = (k+1)/2$, $q_b = u_1/2v_1 - u_2/2v_2$,  and $I_b$ is defined as
\vspace{-1mm}
\begin{eqnarray}\label{eqn:I_b}
	I_b = \frac{v_2^{(j+1)/2}}{2}  \begin{cases}
	\Gamma\left(\frac{j+1}{2}, \frac{u_2^2}{4v_2}\right),  & \text{for} \,\, \frac{u_2}{2v_2} > 0,  \\
	(-1)^j \gamma\left(\frac{j+1}{2}, \frac{u_2^2}{4v_2}\right)
	+ \Gamma\left(\frac{j+1}{2}\right),& \text{otherwise} .
	\end{cases}
\end{eqnarray} 

\vspace{-1mm}
\noindent Moreover, $B$, $s$, $r$, and $\{u_i,v_i\}$ for $i \in \{1,2\}$  are defined as follows:
\vspace{-1mm}
\begin{subequations}\label{eqn:definitions}
\begin{eqnarray} 
	B &=& \frac{m_u^{m_u} \psi \Exp{\frac{u_1^2}{4v_1} -\frac{\mu_{Y}^2}{2\sigma_{Y}^2} - \frac{1}{4a_1}\left(d\vartheta_b - \frac{\mu_{Y}}{\sigma_{Y}}\right)^2 }}{2 \Gamma(m_u) \xi_u^{m_u} \sqrt{2\pi \sigma_{Y}^2 a_1}}, \label{eqn:def_B} \quad \text{and} \quad 
	s = \frac{c\vartheta_b^2}{2\sqrt{a_1}}, \label{eqn:def_s} \\
 	r &=& \frac{1}{2\sqrt{a_1}} \left(d \vartheta_b - \frac{\mu_{Y}}{\sigma_{Y}^2}\right), \label{eqn:def_r} \quad \text{and} \quad
	u_1 = d \vartheta_b - \frac{c\vartheta_b^2}{a1}\left(d\vartheta_b - \frac{\mu_{Y}}{\sigma_{Y}^2}\right), \label{eqn:def_u1}  \\
	u_2 &=& u_1 + 2sr, \label{eqn:def_u2} \quad \text{and} \quad
	v_1 = c \vartheta_b -\frac{c^2 \vartheta_b^2}{a_1} + \frac{m_u}{\xi_u}, \label{eqn:def_v1} \quad \text{and} \quad
 	v_2 = v_1 + s^2, \label{eqn:def_v2}
\end{eqnarray}  		 
\end{subequations}

\vspace{-1mm}
\noindent where $a_1 = c\vartheta_b^2 + 1/2\sigma_{Y}^2$, $c=0.374$, and $d = 0.777$ \cite{Andrew2013}.

\subsection{Achievable Diversity Order }\label{sec:diversity}

The diversity order is  the negative slope of the average SER or   outage probability   versus the average SNR curve    in a log-log scale in the   high SNR regime   \cite{Wang2003a}, and it can be defined as 
\vspace{-1mm}
\begin{eqnarray}\label{eqn:diversiy_def}
	G_d = - \lim_{\bar{\gamma} \rightarrow \infty} \frac{\log{\bar{P}_e}}{\log{\bar{\gamma}}} = - \lim_{\bar{\gamma} \rightarrow \infty} \frac{\log{P_{out}}}{\log{\bar{\gamma}}}.
\end{eqnarray} 

\vspace{-1mm}
\noindent The diversity order provides insights on how the outage probability or average SER decays with the average SNR. Thus,  we derive the asymptotic outage probability and  the average SER in the high SNR regime by using the first order polynomial expansion of the exact PDF of the corresponding RVs.  The achievable diversity order and array/coding gain are also quantified.   
It is worth noting that our high SNR performance metrics are asymptotically accurate as we have not used   CLT approximation for the corresponding derivations (see Appendix \ref{app:Appendix5}). Thus, the achievable diversity order and array gains are accurate,  and they are not approximated performance metrics.

\subsubsection{Asymptotic outage probability}\label{sec:asymp_outage}

 The outage probability can be asymptotically approximated in the high SNR regime  as  \cite{Wang2003a}
 \vspace{-1mm}
\begin{eqnarray}\label{eqn:asym_outage_def}
	\lim_{\bar{\gamma} \rightarrow \infty} {P}_{out} = {P}^{\infty}_{out} \approx O_c \left( \frac{\gamma_{th}}{\bar{\gamma}}\right)^{G_d}+ \mathcal{O} \left(\bar{\gamma}^{-(G_d+1)}\right),
\end{eqnarray}

\vspace{-1mm}
\noindent where $O_c$ is a measure of the array/coding gain and $G_d$ is the diversity order \cite{Wang2003a}. Then, an asymptotic approximation for  $P_{out}$ can be derived as (see Appendix \ref{app:Appendix5})
\vspace{-1mm}
\begin{eqnarray}\label{eqn:asym_outage}
 	{P}^{\infty}_{out} =  \Omega \Phi(N,L)   \left(\frac{\gamma_{th}}{\bar{\gamma}}\right)^{G_d} + \mathcal{O} \left(\bar{\gamma}^{-(G_d+1)}\right),
\end{eqnarray}

\vspace{-1mm}
\noindent where  the diversity order ($G_d$) is given by 
\vspace{-3mm}
\begin{eqnarray}\label{eqn:Gd}
 	G_d =   NL \min{(m_h,m_g)}+m_u.
\end{eqnarray}

\vspace{-2mm}
\noindent Moreover, in \eqref{eqn:asym_outage}, $\Omega$ and $ \Phi(N,L)$ are defined as 
\vspace{-1mm}
\begin{eqnarray}\label{eqn:omega}
 	\Omega =  \frac{ m_u^{m_u} \Gamma(2m_u) }{G_d \Gamma(m_u)\xi_u^{m_u} \Gamma(2G_d) },\qquad \text{and}\qquad
 	\Phi(N,L) =  \prod\nolimits_{n=1}^{N} \prod\nolimits_{l=1}^{L} \eta_{nl} \theta_{n}, \label{eqn:phi}
\end{eqnarray} 

\vspace{-1mm}
\noindent where $m_l = \max{(m_h,m_g)}$, $m_s = \min{(m_h,m_g)}$, and
\vspace{-1mm}
\begin{subequations}
\begin{eqnarray} 
	\theta_{n} &=&  \alpha' \sqrt{\pi} \left(4 \sqrt{\frac{m_s m_l}{\xi_{s_n} \xi_{l_n}}}\right)^{(m_s-m_l)} \begin{cases}
	\frac{|\ln{(\epsilon)}| \Gamma(2m_s+1/2)}{ \sqrt{\pi}\Gamma(2m_s) },  & \text{for} \,\, m_s=m_l,  \\
	\frac{\Gamma(2m_s) \Gamma(2m_l-2m_s)}{\Gamma(m_l-m_s+1/2)},& \text{otherwise},
	\end{cases} \label{eqn:theta_n} \\ 
	\alpha' &=&  \frac{4 }{ \Gamma(m_s) \Gamma(m_l)} \left(\frac{m_s m_l}{\xi_{s_n} \xi_{l_n}}\right)^{(m_s+m_l)/2}. \label{eqn:alpha_dash} 
\end{eqnarray}  		 
\end{subequations}
 
 \vspace{-1mm}
 \noindent Furthermore,  $\xi_{s_n}$ and $\xi_{l_n}$ are the scaling parameters related to $m_s$ and $m_l$, respectively, in the Nakagami-$m$ fading channels, and $1\ll\epsilon<0$ such that $\lim_{z \rightarrow 1^{-}} 1-z$.
 Finally, $O_c$ in (\ref{eqn:asym_outage_def}) can be defined as
 \vspace{-2mm}
\begin{eqnarray}
	O_c =  \Omega \Phi(N,L).
\end{eqnarray}

\vspace{-2mm}

\subsubsection{Asymptotic average SER}\label{sec:asymp_ASER}
Similarly, the average SER can be approximated in the high SNR regime as \cite{Wang2003a}
\vspace{-1mm}
\begin{eqnarray}\label{eqn:asym_ber_def}
 	\lim_{\bar{\gamma} \rightarrow \infty} \bar{P}_e = \bar{P}^{\infty}_e \approx \left(G_a \bar{\gamma}\right)^{-G_d}+ \mathcal{O} \left(\bar{\gamma}^{-(G_d+1)}\right),
\end{eqnarray} 

\vspace{-1mm}
\noindent where $G_d$ and $G_a$ are diversity order and array gain, respectively. 
 An asymptotic approximation for the average SER    \eqref{eqn:ber_alt} in high SNR regime can be derived as (see Appendix \ref{app:Appendix5})
 \vspace{-1mm}
\begin{eqnarray}\label{eqn:asym_ber}
 	\bar{P}^{\infty}_e &=& \left[(\Lambda \Phi(N,L))^{-\frac{1}{G_d}} \bar{\gamma}\right]^{-G_d} + \mathcal{O} \left(\bar{\gamma}^{-(G_d+1)}\right),
\end{eqnarray} 

\vspace{-1mm}
\noindent where $G_d = m_s NL+m_u$ and $\Phi(N,L)$ is given in \eqref{eqn:phi}. Moreover, $\Lambda$ is defined as
\vspace{-1mm}
\begin{eqnarray}\label{eqn:lambda}
	\Lambda &=& \frac{\omega 2^{G_d-1} m_u^{m_u}}{ \sqrt{\pi} \vartheta^{G_d} G_d \xi_u^{m_u}  } \frac{\Gamma(G_d+1/2) \Gamma(2m_u)}{\Gamma(2G_d) \Gamma(m_u)}.
\end{eqnarray} 

\vspace{-1mm}
\noindent Lastly, the array gain in (\ref{eqn:asym_ber_def}) can be derived as
\vspace{-1mm}
\begin{eqnarray}
	G_a = (\Lambda \Phi(N,L))^{-{1}/{G_d}}.
\end{eqnarray}

\vspace{-1mm}


\noindent \textbf{\textit{Remark 5:}} 
Our asymptotic performance analysis (\ref{eqn:Gd}) reveals that the   diversity order is a linear function of the number of distributed IRSs, the number of passive reflective elements in each IRS, and the $m$ parameters of the Nakagami fading channels. It is worth noting that $S$ and $D$ are equipped with a single antenna/RF chain. However, as per  analysis, we observe that the distributed IRSs deployment can be used to recycle the  existing EM waves in the propagation environment without using additional active RF chains  to substantially increase the diversity order, and consequently, the system reliability can be boosted in terms of lowering the outage probability and average SER. It is worth noting that  the current state-of-the-art wireless systems,  the diversity order improvements are obtained by virtue of increasing active RF chains and antennas at the transmitter/receiver. 

\noindent \textbf{\textit{Remark 6:}} 
By varying  $m$ parameter of Nakagami fading, a myriad of fading environments can
be investigated. For instance, higher the $m$ value, lesser the fading severity. The  diversity order of a single reflected path corresponding to an IRS depends on the minimum value of Nakagami-$m$ parameter of $S$-to-IRS and IRS-to-$D$ channels. 
Thus,   more severe channel (or equivalently, the channel with lesser $m$ value) always determines the  diversity order of a single IRS-aided reflected channel. However, when there exist distributed IRSs and a direct channel, the overall diversity order is the sum of the diversity orders of each end-to-end serviceable channel as per \eqref{eqn:Gd}.
The proposed distributed IRSs-aided system achieves this  diversity gain by intelligently controlling the
phase-shifts at the distributed passive reflective elements such that the  signals received via reflected channels are constructively combined with the direct channel at $D$.

\section{The Detrimental Effects of Transmission Impairments}\label{sec:transmission_imp}
\subsection{Effect of Phase Quantization Errors}\label{sec:phase_estimation}
Continuous phase-shift  adjustments at the passive reflecting elements of the IRSs can be  practically infeasible due to hardware  limitations. Thus, we investigate the effect of erroneous phase-shifts at the IRSs by adopting a discrete phase-shift model to capture  quantized phases.  To this end, the IRS controller is allowed to select a discrete phase-shift for each reflecting element from a limited number of quantized phases. Thus, the selected discrete quantized phase for the $l$th reflective element in the $n$th IRS can be defined as
\vspace{-1mm}
\begin{eqnarray}\label{eqn:est_phase}
\hat{\theta}_{nl}^*= \pi \varsigma/2^{B-1}, \quad \text{for} \quad l \in \{1,\cdots,L\} \quad \text{and} \quad n \in \{1,\cdots,N\},
\end{eqnarray}     

\vspace{-1mm}
\noindent where $B$ is the number of quantization bits. Moreover, $\varsigma$ is defined as
\vspace{-1mm}
\begin{eqnarray}\label{eqn:def_varsigma}
\varsigma =\underset{q\in \{0,\pm 1, \cdots, \pm 2^{B-1} \} }{ \mathrm{argmax}} |{\theta}_{nl}^* - \pi q/2^{B-1}| ,
\end{eqnarray}

\vspace{-1mm}
\noindent where $\theta_{nl}^*$ is  optimal phase-shift given in \eqref{eqn:opt_theta}. Then, the error in continuous and quantized phase-shifts can be defined as
$\varepsilon_{nl} = {\theta}_{nl}^*-\hat{\theta}_{nl}^*$, 
which can shown to be uniformly distributed in the regime of large quantization levels as $\varepsilon_n \sim \mathcal{U} \left[-\tau,\tau \right) $, where  $\tau=\pi/2^B$ \cite{Haykin2009}. Thus, we   rewrite the optimal SNR given in  \eqref{eqn:snr_opt} with quantized phase-shifts as follows:
\vspace{-1mm}
\begin{eqnarray}\label{eqn:snr_opt_q}
\gamma^*= \bar{\gamma} \left|\alpha_u  + \sum\nolimits_{n=1}^{N} \sum\nolimits_{l=1}^{L} \eta_{nl} \alpha_{g_{nl}} \alpha_{h_{nl}} \exp{j \varepsilon_{nl}}  \right|^2 = \bar{\gamma} \left[\left(\alpha_u+ Y_R\right)^2+ Y_I^2\right],
\end{eqnarray} 

\vspace{-1mm}
\noindent where $Y_R = \sum_{n=1}^{N} \sum_{l=1}^{L} \eta_{nl} \alpha_{{g}_{nl}}  \alpha_{{h}_{nl}} \cos(\varepsilon_{nl})$ and $Y_I = \sum_{n=1}^{N} \sum_{l=1}^{L} \eta_{nl} \alpha_{{g}_{nl}}  \alpha_{{h}_{nl}} \sin(\varepsilon_{nl})$. Then, we can derive an upper bound for the average achievable rate with   quantized phase-shifts by following the derivation steps similar to those in Appendix \ref{app:Appendix31} as 
\vspace{-1mm}
\begin{eqnarray} \label{eqn:rate_ub_sub_q}
\hat{\mathcal{R}}_{ub} =   \log[2]{1+ \bar{\gamma} \left(\sigma_{u}^2 + \sigma_{Y_R}^2 + \sigma_{Y_I}^2+ 2\mu_{u} \mu_{Y_R} + \mu_{u}^2 + \mu_{Y_R}^2 \right)},
\end{eqnarray}

\vspace{-1mm}
\noindent where $\mu_{u}$ and $\sigma_{u}^2$ are given in \eqref{eqn:u_mean} and \eqref{eqn:u_var}, respectively. Moreover, $\mu_{Y_R}$, $\sigma_{Y_R}^2$, and $\sigma_{Y_I}^2$ are defined as
\vspace{-1mm}
\begin{subequations}
	\begin{eqnarray}\label{eqn:def_YR_YI}
	\!\!\!\!\!\!\!\!\!\!\!\mu_{Y_R} &=& \sum_{n=1}^{N} \sum_{l=1}^{L} \eta_{nl}  \sqrt{\frac{\xi_{h_n} \xi_{g_n}}{m_h m_g}} \frac{\Gamma\left(m_h + 1/2\right) \Gamma\left(m_g+1/2\right)}{\Gamma\left(m_h\right) \Gamma\left(m_g\right)}  \frac{\sin{(\tau)}}{\tau} , \label{eqn:mu_YR} \\
	\!\!\!\!\!\!\!\!\!\!\!\sigma_{Y_R}^2 \! &=& \!\sum_{n=1}^{N} \! \sum_{l=1}^{L}\! \eta_{nl}^2 {\frac{\xi_{h_n} \xi_{g_n}}{m_h m_g}} \!\! \left[ \!\frac{\Gamma\!\left(\! m_h \!+\! 1\right) \! \Gamma\!\left(\!m_g\!+\!1\right)}{\Gamma\left(m_h\right) \Gamma\left(m_g\right)} \!\left(\! \frac{\sin(2\tau)}{4 \tau} \!+\! \frac{1}{2} \!\right)  \!\!-\!\! \left( \!\!\frac{\Gamma\!\left(\!m_h \!+\! \frac{1}{2}\right) \! \Gamma\!\left(\!m_g\!+\!\frac{1}{2}\right)}{\Gamma\left(m_h\right) \Gamma\left(m_g\right)} \! \frac{\sin{(\tau)}}{\tau}\! \right)^{\!\!2} \!  \right]\!\! , \label{eqn:sigma_YR} \\
	\!\!\!\!\!\!\!\!\!\!\!\sigma_{Y_I}^2 &=& \sum_{n=1}^{N} \sum_{l=1}^{L} \eta_{nl}^2 {\frac{\xi_{h_n} \xi_{g_n}}{m_h m_g}} \frac{\Gamma\left(m_h + 1\right) \Gamma\left(m_g+1\right)}{\Gamma\left(m_h\right) \Gamma\left(m_g\right)} \left(\frac{1}{2} - \frac{\sin(2\tau)}{4 \tau} \right). \label{eqn:sigma_YI}
	\end{eqnarray} 
\end{subequations}

	\noindent \textbf{\textit{Remark 7:}} 
	In an IRS-aided communication system, the imperfectly estimated cascaded channels   result in imperfect  phase-shift adjustments at the reflective elements \cite{Badiu2020}. Thus, the effects of imperfectly estimated channels can be modeled as phase-shift errors, which are randomly distributed on $\left[-\pi,\pi \right)$ based on a certain circular distribution  \cite{Badiu2020}. Hence, our analysis on phase quantization errors can be extended to capture the impacts of imperfect channel estimation of the proposed distributed IRS-aided wireless system.

\subsection{Effect of Spatially-Correlated Fading}\label{sec:correlation}

To capture deleterious effect of correlated fading for the reflecting elements at the IRSs,  we consider a two-dimensional rectangular IRS having $L=L_H \times L_V$  elements, where $L_H$ and $L_V$ are the numbers of elements per row  and column, respectively \cite{Renzo2020a}. Moreover, we assume that each reflective element has an area of $A=d_H\times d_V$, where $d_H$ is the horizontal width and $d_V$ is the vertical height. Then, the correlation matrix at the $n$th IRS can be modeled as $\mathbf{R}_{v_n} \in \mathbb C^{L\times L}$ for $v \in \{h,g\}$, where the $(i,j)$th element of $\mathbf{R}_{v_n}$ is given by \cite{Emil2020a}
\vspace{-1mm}
\begin{eqnarray}\label{eqn:ij_elemnt}
[\mathbf{R}_{v_n}]_{i,j} = A \sinc{\left(2\|\mathbf{u}_{v_n,i}-\mathbf{u}_{v_n,j} \|/\lambda\right)},
\end{eqnarray}  

\vspace{-1mm}
\noindent where $\lambda$ is the wavelength of the plane wave, and $\mathbf{u}_{v_n,k}$ for $k\in \{i,j\}$ is defined as
\vspace{-1mm}
\begin{eqnarray}\label{eqn:u_def}
\mathbf{u}_{v_n,k} = [0, \mod{\!\!\left(k-1,L_H\right)}d_H, \quad \lfloor(k-1)/L_H\rfloor d_V]^T.
\end{eqnarray}  

\vspace{-1mm}
\noindent 
When there exists correlated fading,  $S$-$n$th IRS and $n$th IRS-$D$ channels can be modeled as
\vspace{-1mm}
\begin{eqnarray}\label{eqn:cor_channels}
\tilde{\mathbf{h}}_n = \mathbf{R}_{h_n}^{1/2} \mathbf{h}_n, \qquad \text{and} \qquad \tilde{\mathbf{g}}_n^T = \mathbf{g}_n^T \mathbf{R}_{g_n}^{1/2},
\end{eqnarray} 

\vspace{-1mm}
\noindent where $\mathbf{h}_n = [h_{n1},\cdots, h_{nl}, \cdots, h_{nL}]^T\in \mathbb C^{L\times 1}$  and  $\mathbf{g}^T_n = [g_{n1},\cdots, g_{nl}, \cdots, g_{nL}]\in \mathbb C^{1\times L}$. Then, the SNR at $D$ can be written as
\vspace{-1mm}
\begin{eqnarray}\label{eqn:snr_cor}
\gamma &=& \bar{\gamma} \left|u + \sum\nolimits_{n=1}^{N} \tilde{\mathbf{g}}^T_n \mathbf \Theta_n \tilde{\mathbf{h}}_n \right|^2 = \bar{\gamma} \left|\alpha_u \exp{j \theta_{u}} + \sum\nolimits_{n=1}^{N} \mathbf{g}_n^T \mathbf{R}_{g_n}^{1/2} \mathbf \Theta_n \mathbf{R}_{h_n}^{1/2} \mathbf{h}_n \right|^2.
\end{eqnarray}

\vspace{-1mm}
\noindent In the presence of spatially correlated fading, the SNR in \eqref{eqn:snr_cor} can be maximized by considering the phases introduced by correlation matrices. To this end, the  optimal choice of $\theta_{nl}$ to maximize the received SNR at $D$ can be given by
\vspace{-1mm}
\begin{eqnarray}\label{eqn:opt_theta_cor}
\theta_{nl}^* = \theta_u - \left(\tilde{\theta}_{g_{nl}} + \tilde{\theta}_{h_{nl}}\right), \quad \text{for}\quad n \in \{1,\cdots,N\} \quad \text{and} \quad l \in \{1,\cdots,L\},
\end{eqnarray} 

\vspace{-1mm}
\noindent where $\tilde{\theta}_{h_{nl}}$ and $\tilde{\theta}_{g_{nl}}$ are phases of $[\tilde{\mathbf{h}}_n]_{l,1}$ and $[\tilde{\mathbf{g}}_n^T]_{1,l}$, respectively, which account for the cumulative phases of the corresponding correlation matrices and independently faded channels. Thereby, the     average achievable rate for the distributed IRS set-up with correlated fading  can be written as
\vspace{-1mm}
\begin{eqnarray}\label{eqn:snr_cor_opt}
 \mathcal{R} =\E{\log[2]{1+ \bar{\gamma} \left|\alpha_u + \sum\nolimits_{n=1}^{N} \hat{\mathbf{g}}^T_n \boldsymbol \eta_n \hat{\mathbf{h}}_n \right|^2}},
\end{eqnarray}

\vspace{-1mm}
\noindent where $\boldsymbol \eta_n= \diag{\eta_{n1} , \cdots, \eta_{nl} , \cdots, \eta_{nL} }\in \mathbb C^{L\times L}$. Moreover, $\hat{\mathbf{h}}_n$ and $\hat{\mathbf{g}}_n$ represent  the vectors with the moduli of the elements of $\tilde{\mathbf{h}}_n$ and $\tilde{\mathbf{g}}_n$, respectively.

\begin{figure}[!t]\centering \vspace{-7mm}
	\def\svgwidth{230pt} 
	\fontsize{10}{8}\selectfont 
	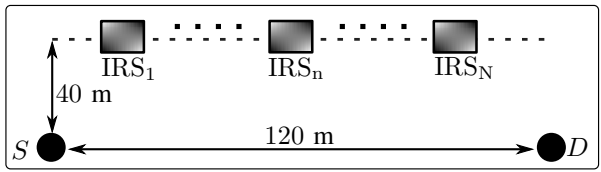 \vspace{-4mm}
	\caption{The simulation set-up of the distributed IRSs-aided communication system. }\label{fig:sim_setup} \vspace{-0mm} 
\end{figure}

\section{Numerical Results}\label{sec:Numerical}

In this section, we present our numerical results to investigate the performance gains and to validate our analysis. We consider the simulation set-up in Fig. \ref{fig:sim_setup}, where $S$ and $D$ are in positioned at fixed locations and  $120$\,m apart from each other. Unless otherwise specified, the IRSs are uniformly deployed	in between $S$ and $D$ such that the minimum distance between $S$ or $D$ and the closest IRS is greater the $40$\,m. 
The large-scale fading  is  modeled  as $\zeta_{ab} = \left(d_0/d_{ab}\right)^{\nu} \times 10^{\varphi_{ab}/10}$, where $a \in \{S,n\}$,  $b \in \{n,D\}$, $d_{ab}$ is the distance between nodes $a$ and $b$, $d_0=1$\,m is a reference distance, $\nu=2.8$ is the path-loss exponent, and $10^{\varphi_{ab}/10}$ captures log-normal shadow fading with $\varphi_{ab} \sim \mathcal{N}(0,8)$ \cite{Marzetta2016_Book}. 
Moreover, the amplitude of reflection coefficients $\eta_{nl}$ for $n \in \{1, \cdots, N\}$ and $l \in \{1, \cdots, L\}$ is set to $0.9$, and  unless otherwise specified, the Nakagami-$m$ parameters ($m_u$, $m_h$, and $m_g$) are set to 3.


\begin{figure}[!t]\centering\vspace{-5mm}
	\includegraphics[width=0.45\textwidth]{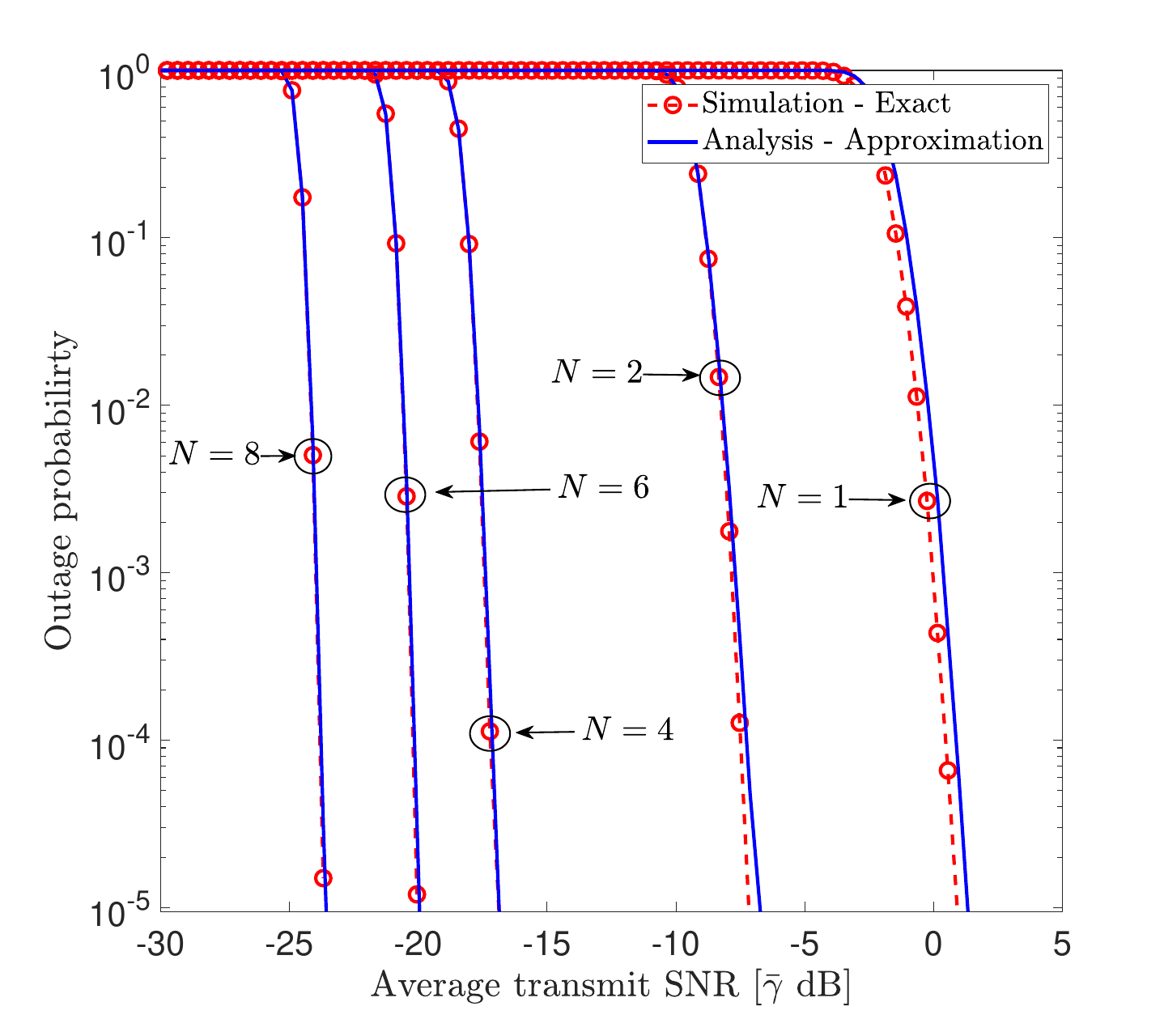}\vspace{-6mm}
	\caption{The outage probability for different number of   IRSs in the distributed deployment denoted by $N \in \{1,2,4,6,8\}$, the number of reflective elements per IRS is $L = 32$, $m_u=m_h=m_g=3$, the threshold SNR is $\gamma_{th} = 0\,$dB.}
	\label{fig:outage_2_4_6_8}\vspace{-5mm}
\end{figure}

In Fig. \ref{fig:outage_2_4_6_8}, we plot the outage probability  as a function of the average transmit SNR ($\bar{\gamma}$) by varying the number of    distributed IRSs defined by  $N \in \{1, 2,4,6,8\}$. The exact outage curves are generated by using  Monte-Carlo simulation, while the analytical outage curves are plotted via our closed-form derivation in \eqref{eqn:out_prob}. As per Fig. \ref{fig:outage_2_4_6_8}, we observe   that the tightness of our  outage probability approximation  significantly improves with $N$. Moreover, we observe that the distributed IRSs deployment outperforms the single-IRS set-up. For example,  the single-IRS deployment  requires an average transmit SNR of $1$\,dB to achieve an outage probability of $10^{-4}$, and this is an $104.2$\% increase compared to the same transmit SNR requirement of the distributed deployment with eight IRSs. 

\begin{figure}[!t]\centering\vspace{-5mm}
	\includegraphics[width=0.45\textwidth]{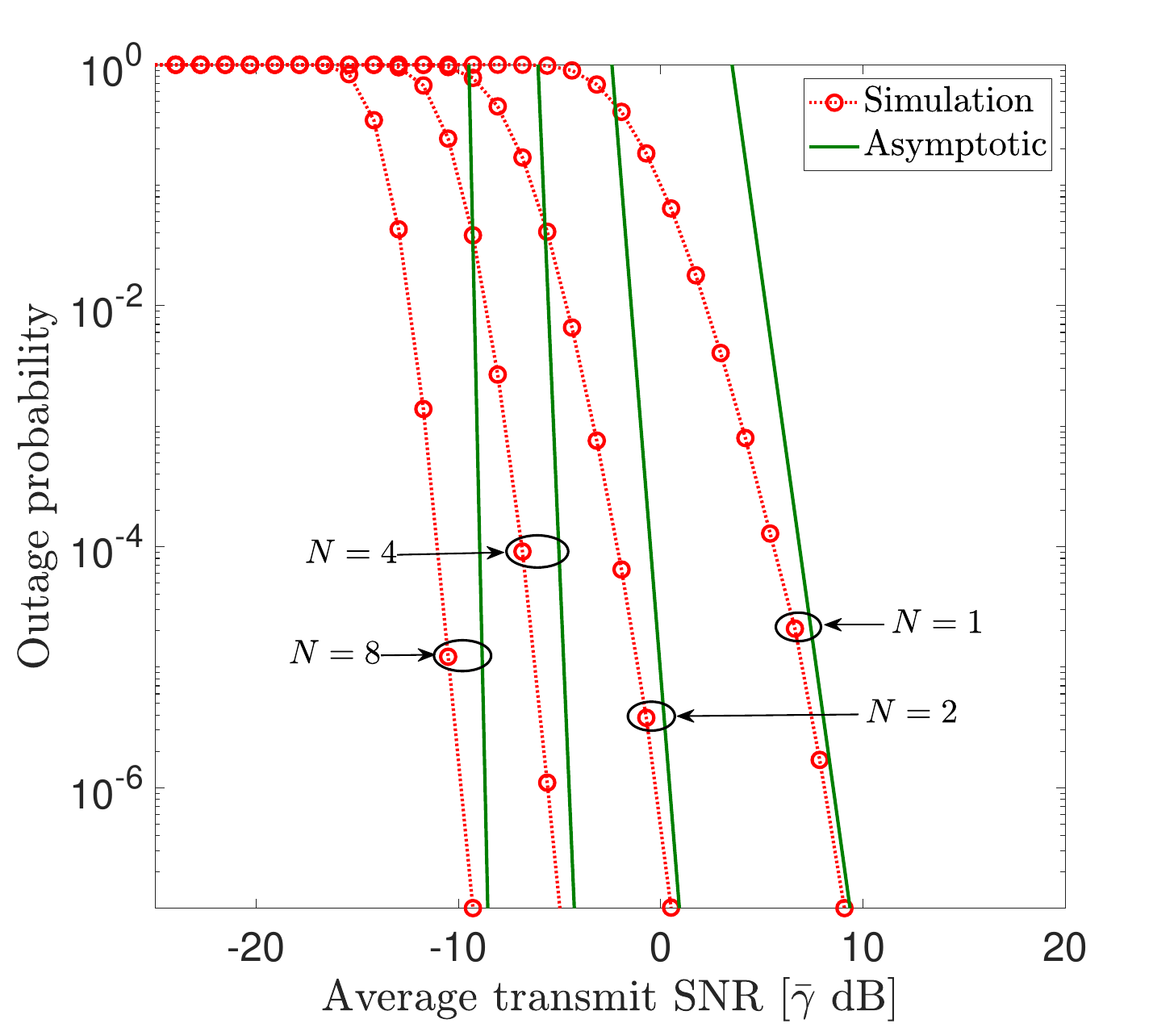}\vspace{-4mm}
	\caption{The outage probability for $N \in \{1,2,4,8\}$, $L = 6$, $m_u =m_g=3$, $m_h=1.5$, and $\gamma_{th} = 10\,$dB.}
	\label{fig:outage_asymp_N_8_4_2_1}\vspace{-5mm}
\end{figure}

In Fig. \ref{fig:outage_asymp_N_8_4_2_1}, we investigate asymptotic behavior of the outage probability in the high SNR regime.  To this end,  our asymptotic outage probability \eqref{eqn:asym_outage} is plotted as a function of the average transmit SNR, and Monte-Carlo simulations are also provided for validation purposes. The asymptotic outage probability curves reveal the achievable diversity orders for different distributed IRSs deployments; $N \in\{1,2,4,8\}$. For instance, the diversity orders for $(N=1, L=6)$ and $(N=2, L=6)$ cases can be obtained by computing the negative gradients of asymptotic outage curves in Fig. \ref{fig:outage_asymp_N_8_4_2_1} to be 12 and 21, respectively. This observation clearly validates our diversity order analysis in \eqref{eqn:Gd}, i.e., $G_d = NL\min(m_h,m_g)+m_u$ with $m_h=1.5$, $m_g=m_u=3$.  
Thus,  the results in  Fig. \ref{fig:outage_asymp_N_8_4_2_1} show that the diversity order increases with the number of IRS reflective elements, and consequently, the system reliability can be drastically improved by the proposed distributed IRSs-aided system  without adopting additional RF chains.  
 

\begin{figure}[!t]\centering\vspace{-5mm}
	\includegraphics[width=0.45\textwidth]{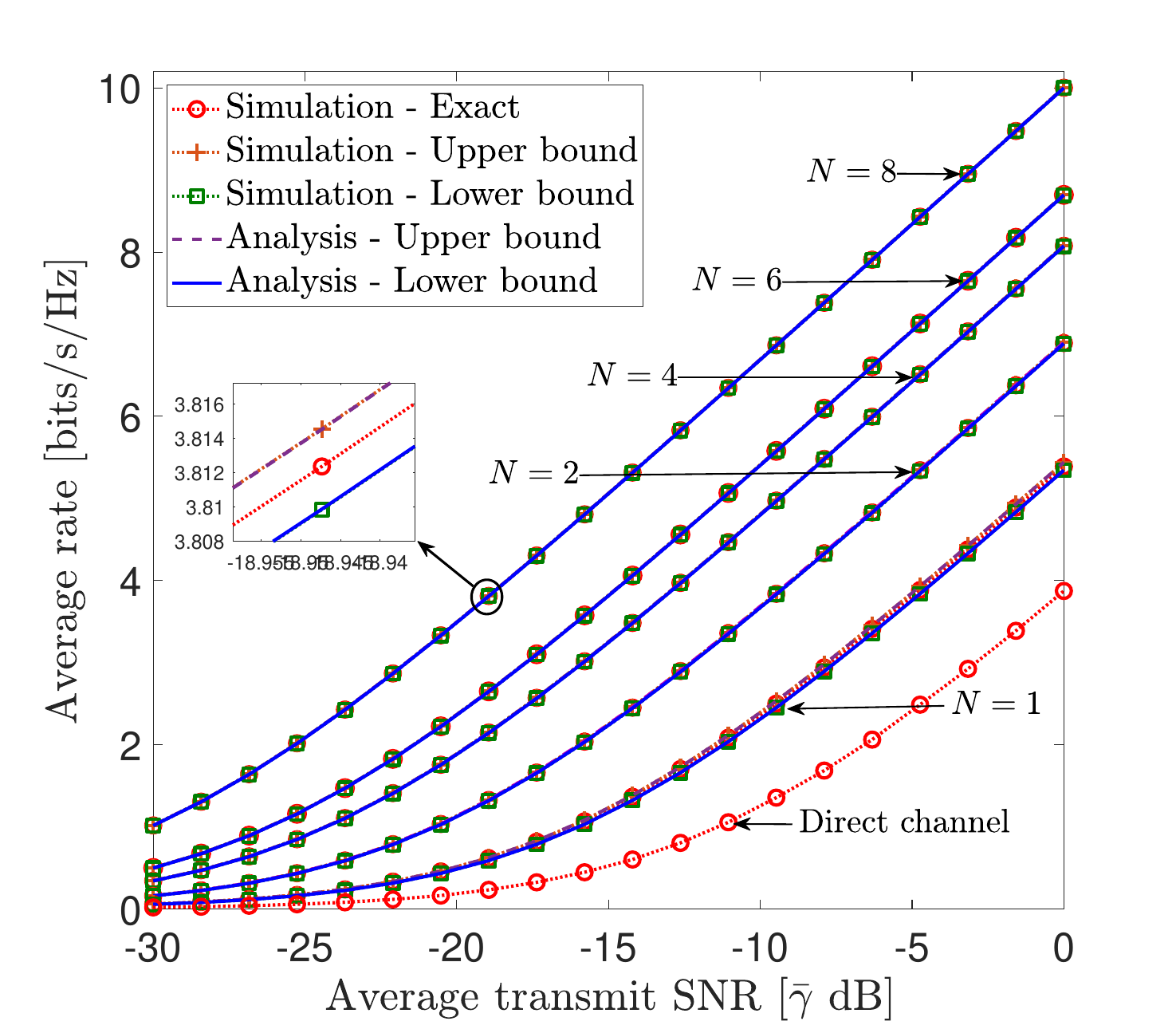}\vspace{-4mm}
	\caption{The average achievable   rate for $N \in \{1,2,4,6,8\}$, $L=32$, and  $m_u=m_h=m_g=3$.}
	\label{fig:rate_N_2_4_6_8}\vspace{-0mm}
\end{figure}

In Fig. \ref{fig:rate_N_2_4_6_8}, we plot the   achievable average rate  for different number of distributed IRSs as per   $N \in \{1,2,4,6,8\}$.  The achievable rate lower/upper bounds are plotted via \eqref{eqn:rate_lb} and \eqref{eqn:rate_ub}, respectively. The accuracy of our rate analysis  is also validated via Monte-Carlo simulation of the exact achievable rate. Fig. \ref{fig:rate_N_2_4_6_8} clearly depicts  that our lower/upper bounds are tight even for  smaller $N$.   Fig. \ref{fig:rate_N_2_4_6_8}   reveals that higher  the number of distributed IRSs, higher the achievable rate. Moreover, Fig.  \ref{fig:rate_N_2_4_6_8} can be used to  quantify the  rate gains of the distributed IRSs deployment compared to the direct transmission. For instance,  the single-IRS ($N=1$) case provides a rate gain of $80.7$\% compared to the direct transmission at an average transmit SNR of $-10$\,dB.  This rate gain increases to $182.5$\%,   $266.2$\%, and $405.9$\% for dual-, quadruple-, and octuple-IRS cases, respectively, with respect to the direct channel.


\begin{figure}[!t]\centering\vspace{-7mm}
	\includegraphics[width=0.45\textwidth]{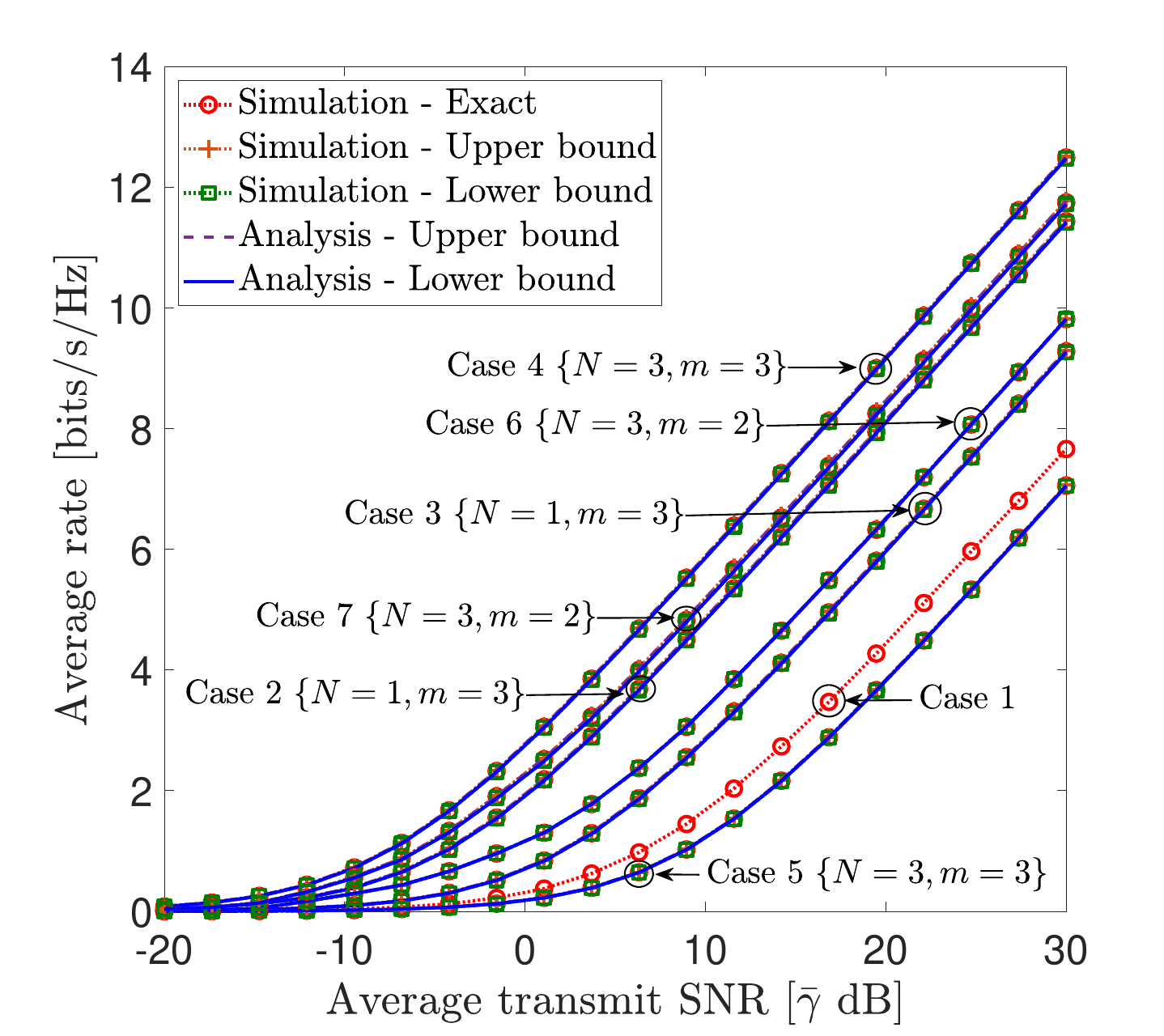}\vspace{-4mm}
	\caption{An average achievable rate comparison for different system configurations. For all cases  $L=32$ and $m_u=m_h=m_g=m$. The cases are defined as follows: Case-1: direct channel only, Case-2/Case-3: IRSs are located in closer with/without direct channel,   Case-4/Case-6/Case-7: IRSs are located  far-away and with direct channel, and Case-5: IRSs are located  far-away and without direct channel, respectively.}
	\label{fig:rate_comparision}\vspace{-5mm}
\end{figure}

In Fig. \ref{fig:rate_comparision}, the effects of the IRS positions and the  shape parameter ($m$) of the Nakagami fading are investigated. To this end, $S$ and $D$ are    positioned at fixed locations at  $50$\,m apart. The  Case-1 plots the achievable  rate of the  direct channel. For Case-2 and Case-3, the distances between $S$-IRS and $D$-IRS are $75$\,m. Moreover,  the distances between $S$-IRS and  $D$-IRS for Case-4 to Case-7 are greater than $160$\,m.  In Case-3 and Case-5, the achievable rates are plotted by assuming that the direct channel is not serviceable, and the end-to-end communication is established via the  reflected channels. Fig. \ref{fig:rate_comparision}  shows that the rate of the direct channel is higher than   that of the reflected channels when the IRSs are positioned far away from $S$ and $D$ (see Case-1 and Case-5). When comparing Case-2 and Case-3, we can observe that   having the direct channel is beneficial in boosting the achievable rate.   Although the IRSs are placed far away in Case-4 than that in Case-3, the achievable rate of the former is higher than that of the latter. This is because Case-4 has two additional IRSs than the Case-3. Thus, the distributed IRSs deployment can circumvent the rate losses incurred  due to larger transmission distances in the reflected channels. Moreover, Case-6 and Case-7 show the  impact of different $m$ parameter of Nakagami fading. In Case-6, $m=1$ leads to have Rayleigh fading channel, whereas  in Case-7, $m=2$ depicts the achievable rate of Rician fading with $1.5$\,dB Rician factor.

%

\begin{figure}[!t]\centering\vspace{-5mm}
	\includegraphics[width=0.45\textwidth]{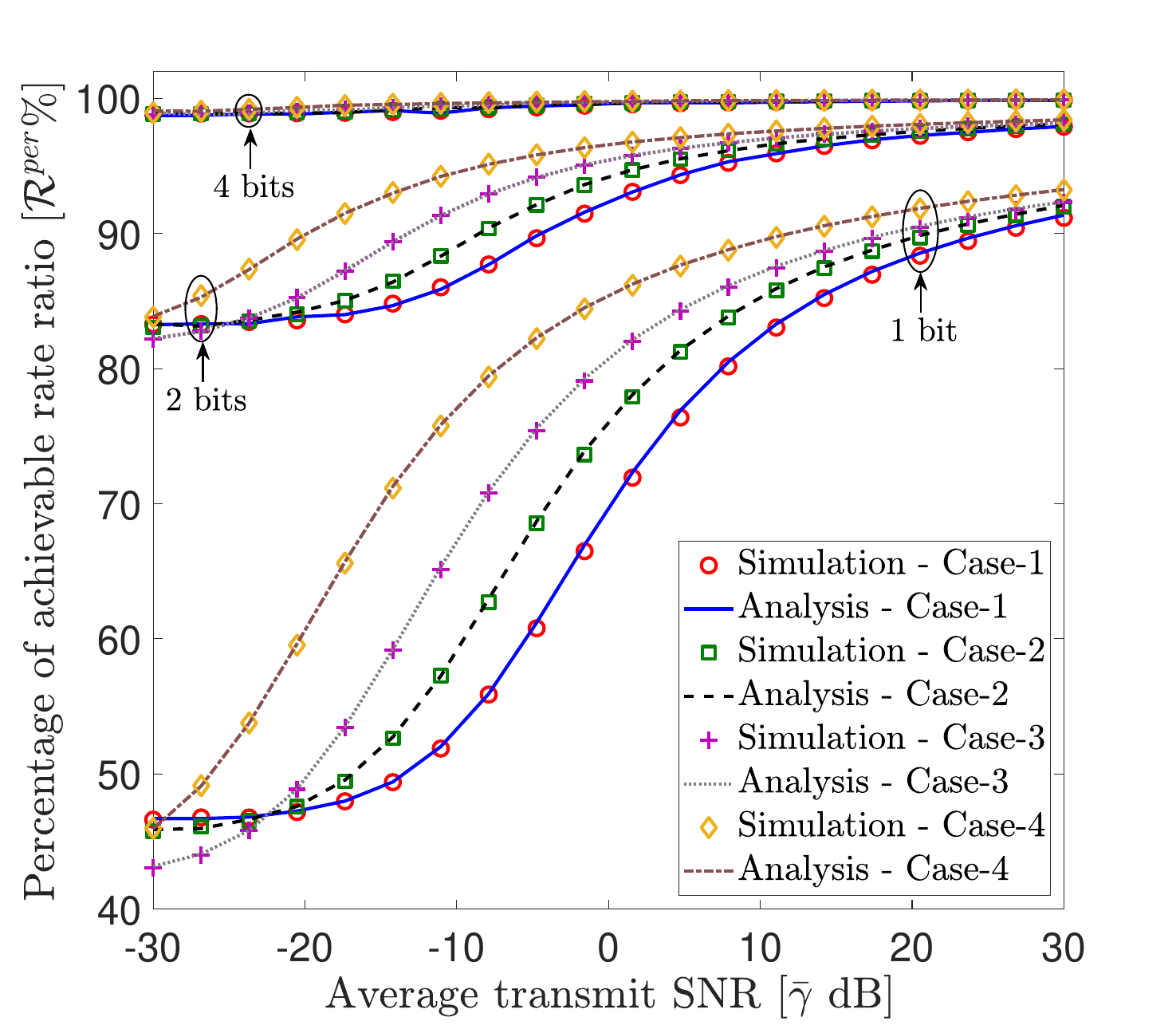}\vspace{-6mm}
	\caption{The impact of phase-shift quantization errors on the achievable rate for  $N \in \{1,2,4,6\}$, $L=32$, and $m_u=m_h=m_g=3$.}
	\label{fig:rate_Q_N_1_2_4_6}\vspace{-5mm}
\end{figure}

In Fig. \ref{fig:rate_Q_N_1_2_4_6}, the effect of   phase-shift quantization errors is studied for different number of IRSs ($N\in \{1,2,4,6\}$) and distinct number of quantization bits $(B\in\{1,2,4\})$, by plotting the percentage rate ratio $(\mathcal{R}^{per})$ as a function of the average transmit SNR. The percentage rate ratio  is defined as $\mathcal{R}^{per}= \hat{\mathcal{R}}_{ub}/{\mathcal{R}}_{ub} \times
100\%$, where ${\mathcal{R}}_{ub}$ denotes the upper bound of the achievable rate with continuous phase-shifts in \eqref{eqn:rate_ub_sub}, while $\hat{\mathcal{R}}_{ub}$ is the average achievable rate upper bound with phase-shift quantization errors   in \eqref{eqn:rate_ub_sub_q}. Monte-Carlo simulations are also plotted   for the purpose of validating the analysis. From Fig. \ref{fig:rate_Q_N_1_2_4_6}, we observe that the effect of  phase-shift quantization errors can be neglected when a higher number of quantization levels is adopted. For example, compared to the continuous phase-shifts,  $99\%$ of the  average achievable rate can be recovered with four-bit phase-shift quantization. Moreover, $\mathcal{R}^{per}$ can be improved by employing a higher number of distributed IRSs. For instance,  at an transmit SNR of $0\,$dB,   $N=2$ case provides a gain of percentage rate ratio of $8.0\%$ compared to the single IRS case for one-bit quantization. Furthermore, this percentage rate gain increases to $13.6\%$ and $19.3\%$ for $N=4$ and $N=6$ cases, respectively.  Fig. \ref{fig:rate_Q_N_1_2_4_6} also depicts that the high transmit SNR is also beneficial for recovering the average achievable rate.

\begin{figure}[!t]\centering\vspace{-5mm}
	\includegraphics[width=0.45\textwidth]{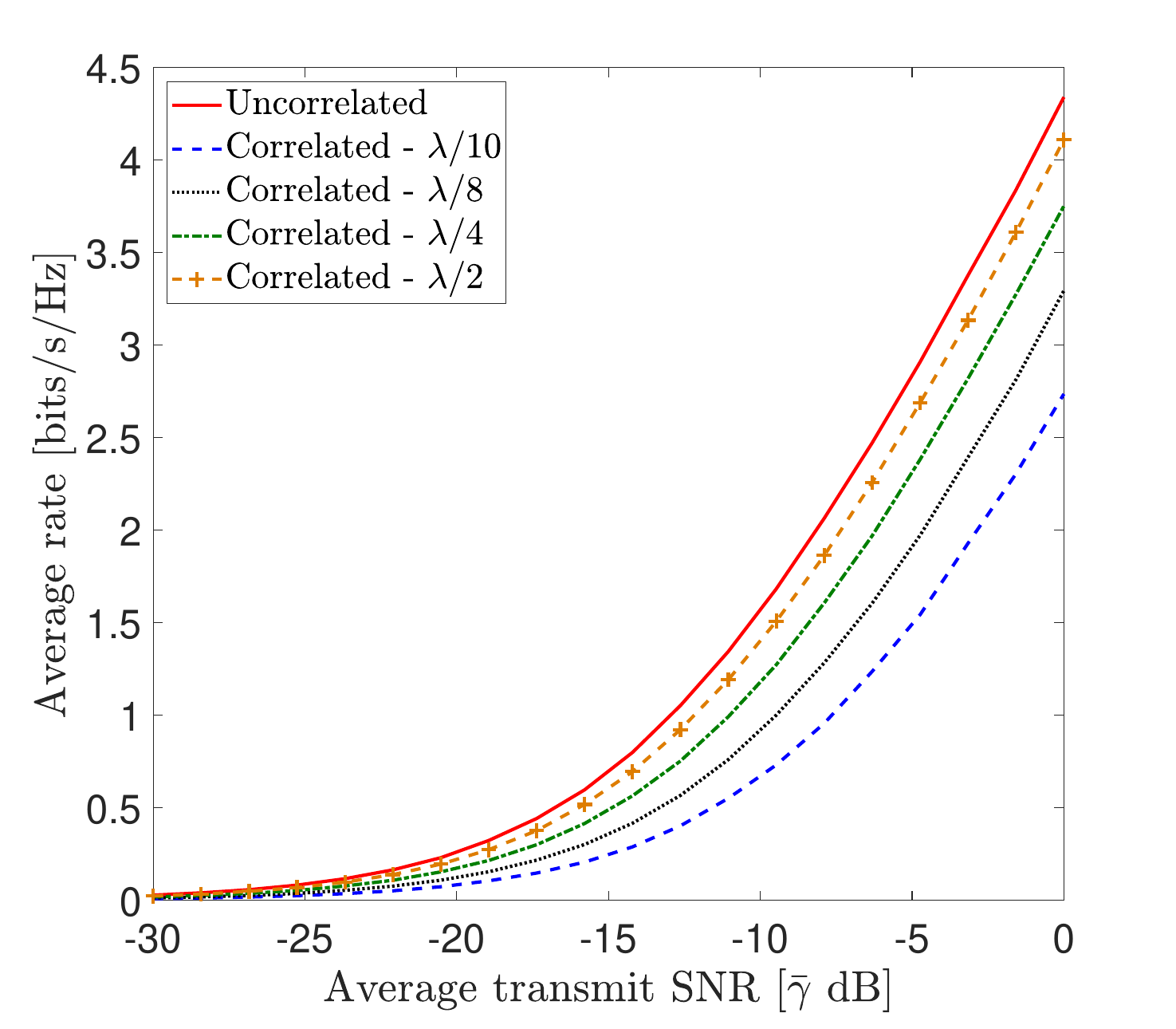}\vspace{-4mm}
	\caption{The impact of correlated fading on the achievable rate for  $N =2$, $L=64$, $m_u=m_h=m_g=3$, and $d_H=d_V \in \{\lambda/2, \lambda/4, \lambda/8, \lambda/10\}$.}
	\label{fig:rate_cor}\vspace{-5mm}
\end{figure}

In Fig. \ref{fig:rate_cor}, the effect of correlated fading is investigated by plotting  the average achievable rate as a function  of the average transmit SNR ($\bar{\gamma}$) for  different sizes of passive reflectors as $d_H=d_V\in\{\lambda/2, \lambda/4, \lambda/8, \lambda/10\}$, respectively. The correlation model in \eqref{eqn:ij_elemnt} is adopted at each IRS for generating the correlated achievable rate curves, while uncorrelated  achievable rate curves are plotted by setting $\mathbf{R}_{v_n}=\mathbf{I}_L$ for $v\in \{h,g\}$ and $n\in \{1,\cdots,N\}$. Fig.  \ref{fig:rate_cor}   clearly shows that the spatially correlated fading  detrimentally affects the achievable rate compared to the uncorrelated fading case. 
Since the correlation matrices are proportional to the area of reflecting elements (i.e., $\mathbf{R}_{v_n} \propto A$),  the achievable rate increases with the size of reflecting elements at IRSs.  
Thus, the performance metrics for the uncorrelated fading  may be used as upper bounds or benchmarks for practical implementations with smaller reflecting elements under spatial correlated fading cases.


\begin{figure}[!t]\centering\vspace{-5mm}
	\includegraphics[width=0.45\textwidth]{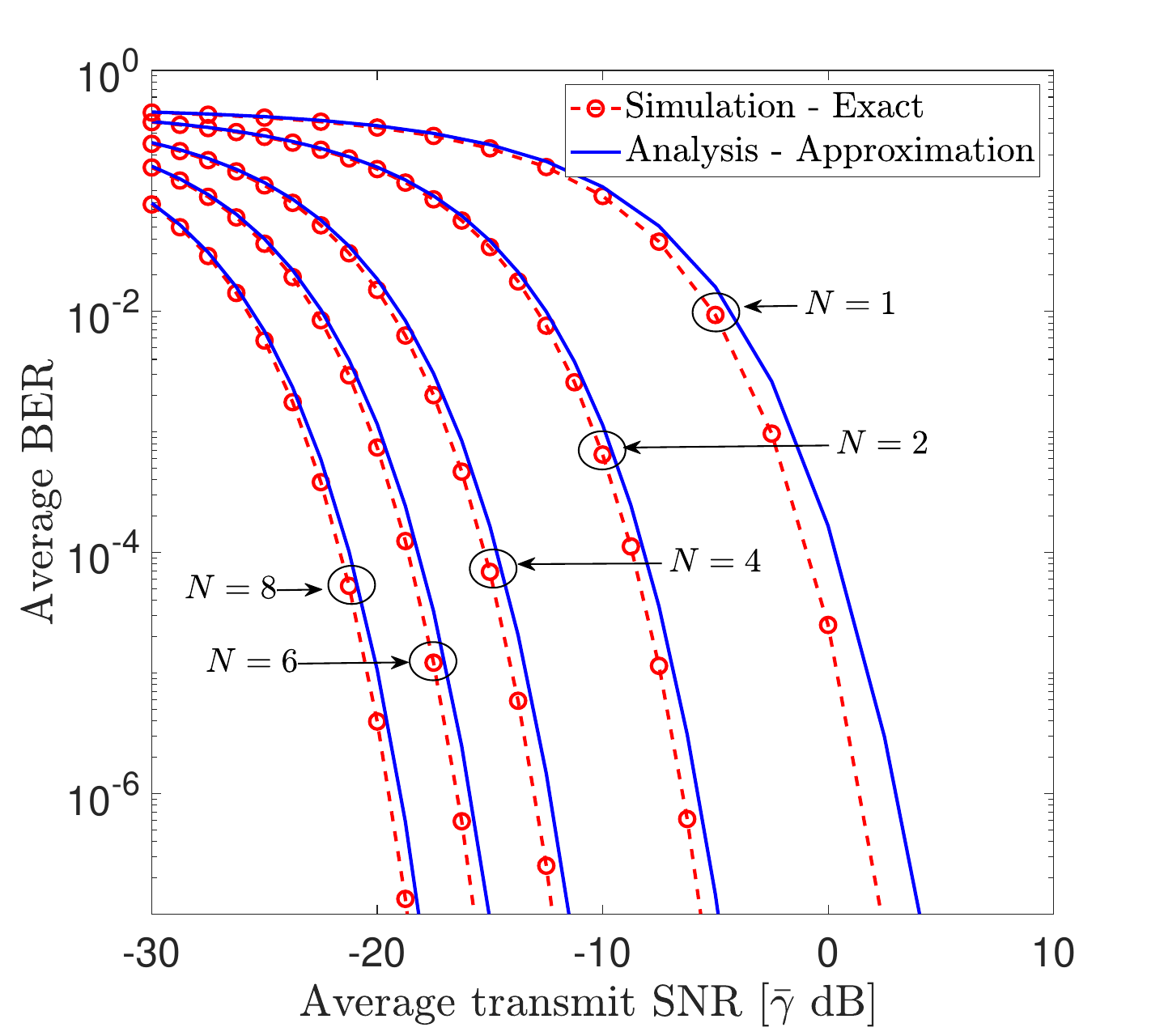}\vspace{-4mm}
	\caption{The average BER of BPSK versus average transmit SNR ($\bar{\gamma}$)   for $N \in \{1,2,4,6,8\}$, $L=32$, $m_u=m_h=m_g=3$, $\omega =  1$,  and $\vartheta = 2$.}
	\label{fig:BER_N_2_4_6_8}\vspace{-5mm}
\end{figure}

In Fig. \ref{fig:BER_N_2_4_6_8}, the average bit error rate (BER) of BPSK  is plotted against the average transmit SNR ($\bar{\gamma}$) for  different number of IRSs;  $N \in \{1,2,4,6,8\}$. The exact average BER curves are generated  via Monte-Carlo simulations, where as the analytical curves are plotted  by using \eqref{eqn:avg_ber}. From Fig.  \ref{fig:BER_N_2_4_6_8}, we observe that our average BER analysis in  \eqref{eqn:avg_ber} is tight for moderately larger $NL$ regime. The average BER for a single-IRS set-up is also plotted for the purpose of comparison. Fig.  \ref{fig:BER_N_2_4_6_8}   reveals  that the distributed IRSs deployment   outperforms the single-IRS set-up. For instance, to achieve an average BER of $10^{-4}$, the single-IRS set-up requires an average transmit SNR of $0$\,dB, which is an $22$\,dB increase compared to the transmit SNR requirement of the distributed IRSs set-up with eight IRSs. Moreover, the achievable average BER   significantly decreases   when the number of distributed IRSs grows large. For example, at an average SNR of $-20$\,dB, the average BER of $N=4$ case is lowered by $99.8$\% by doubling the number of IRSs to $N=8$.

\begin{figure}[!t]\centering\vspace{-5mm}
	\includegraphics[width=0.45\textwidth]{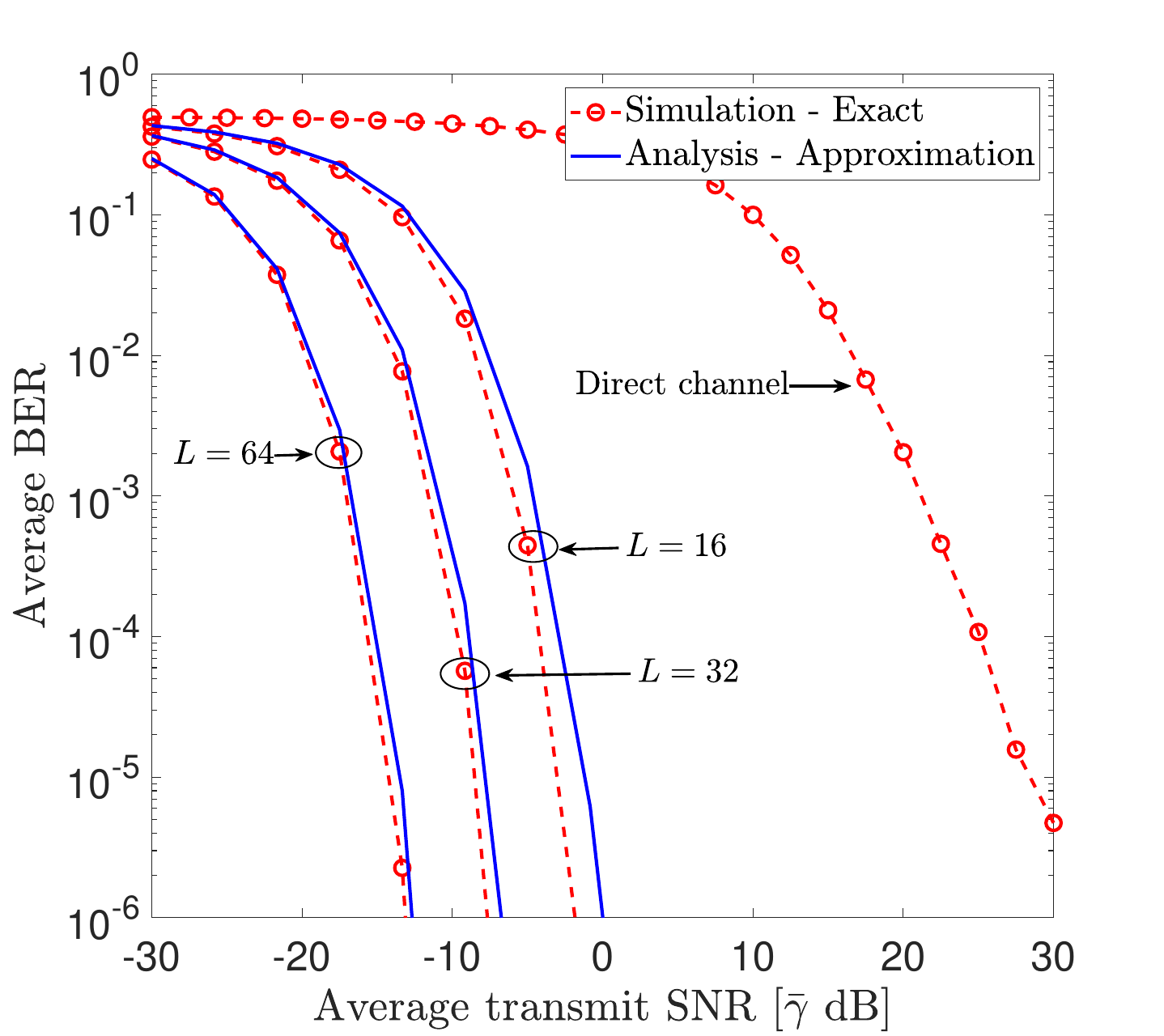}\vspace{-4mm}
	\caption{A comparison of the average BER with/without IRSs for $N =2$, $L \in \{16,32,64\}$, $m_u=m_h=m_g=3$, $\omega =  1$, and $\vartheta = 2$.}
	\label{fig:BER_direct_L_4_8_16_32}\vspace{-7mm}
\end{figure}

In Fig. \ref{fig:BER_direct_L_4_8_16_32}, we investigate the effects of increasing number of IRS reflective elements  by plotting the  average BER of BPSK versus the average SNR.  A dual-IRS system is considered and the number of  reflective elements is varied as  $L \in \{16,32, 64\}$.   Analytical average BER is plotted via \eqref{eqn:avg_ber}, and it is validated through  Monte-Carlo simulations. Fig. \ref{fig:BER_direct_L_4_8_16_32}   shows  that the dual-IRS system outperforms the direct channel in terms of the average BER. For example, the direct transmission requires an average  SNR of $28$\,dB to achieve an average BER of $10^{-5}$, while this SNR requirement can be lowered by $36$\,dB  with a  deployment of a  dual-IRS  set-up with $L=32$.   Moreover, at an average SNR of $-10$\,dB, the average BER can be lowered from $1.82 \times 10^{-2}$ to $1.72 \times 10^{-4}$ by doubling the number of reflective elements per IRS from $L=16$ case to $L=32$ case. Thus, the proposed distributed IRSs deployment can be used to significantly boost the reliability of wireless communication systems.

\begin{figure}[!t]\centering \vspace{-5mm}
	\def\svgwidth{460pt} 
	\fontsize{8}{6}\selectfont 
	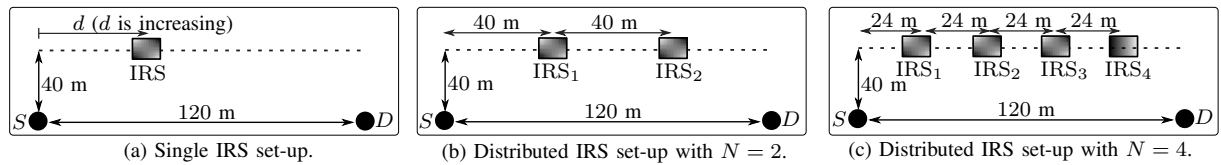 \vspace{-4mm}
	\caption{The simulation set-ups for Fig. \ref{fig:rate_location}. }\label{fig:sim_setup_fig_all} \vspace{-0mm} 
\end{figure}

\begin{figure}[!t]\centering\vspace{-7mm}
	\includegraphics[width=0.45\textwidth]{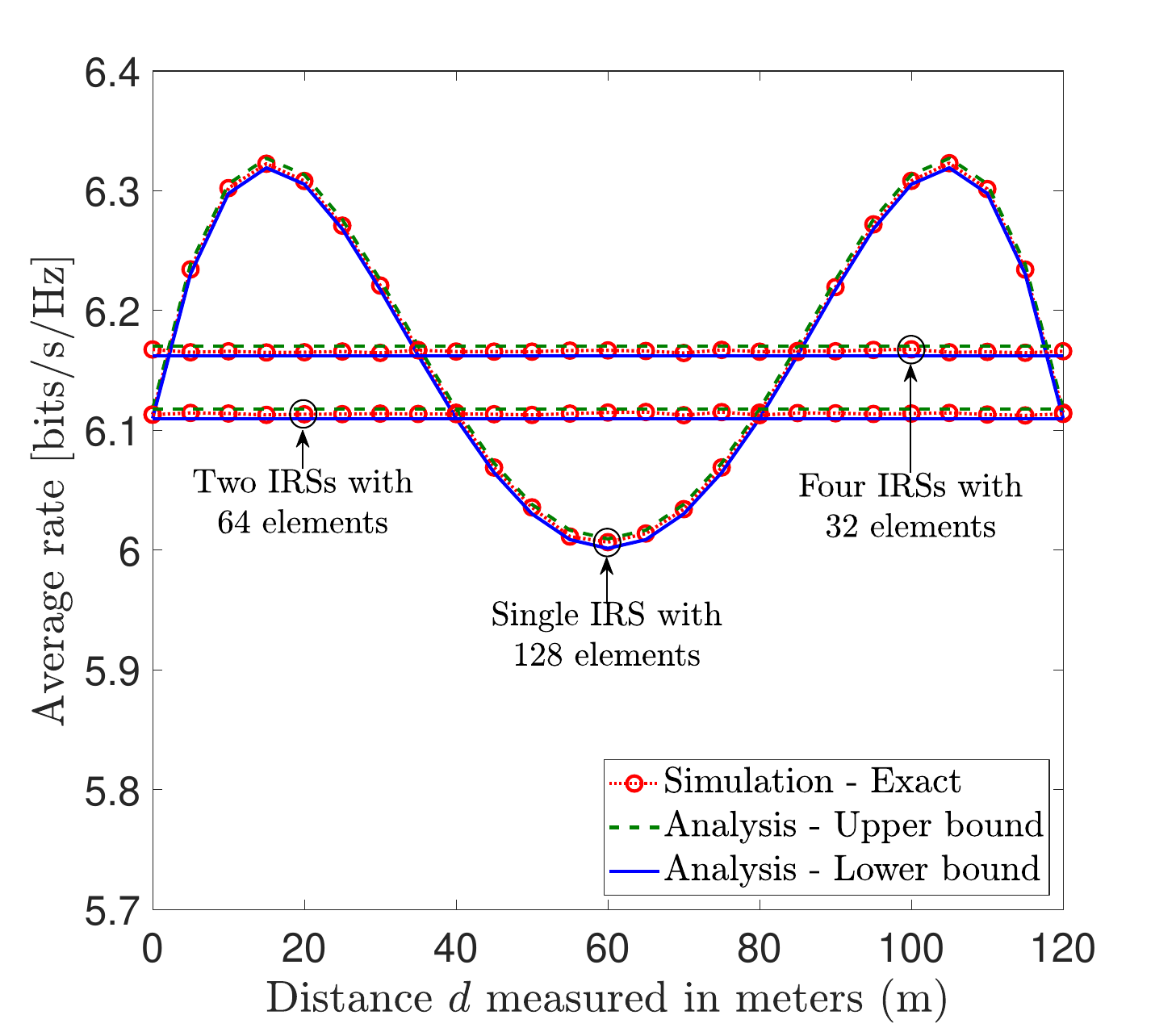}\vspace{-4mm}
	\caption{The average achievable  rate versus the distance for $N \in \{1,2,4\}$, $L=\{128,64,32\}$,   $m_u=m_h=m_g=3$, and $\bar{\gamma}=0$ dB.}
	\label{fig:rate_location}\vspace{-5mm} 
\end{figure}

	In Fig. \ref{fig:rate_location}, we compare the achievable rate performance of a single IRS with $NL$ elements with  $N$ distributed IRSs with $L$ elements. Our aim is to investigate the location regions in which a single IRS would outperform the distributed IRS set-ups    in terms of the achievable rate. To this end, we consider three cases: (a) a single IRS case, where the position of the IRS ($d$) is varied between $S$ and $D$ (see Fig. \ref{fig:sim_setup_fig_all}-a), (b) two uniformly distributed IRSs with $L=64$ (see Fig. \ref{fig:sim_setup_fig_all}-b), and  (c) four uniformly distributed IRSs  with  $L=32$ (see Fig. \ref{fig:sim_setup_fig_all}-c). In the distributed IRS cases, the locations of the IRSs are fixed and for all the cases, the product of the number of IRSs and the number of elements is kept at a fixed value $NL=128$. We plot the average achievable rate with distance $d$.
	The achievable rates of the distributed IRS cases do not vary with $d$ since the corresponding IRSs are kept at fixed locations as shown in Fig. \ref{fig:sim_setup_fig_all}-b and Fig. \ref{fig:sim_setup_fig_all}-c. 
	However, since we change the position of the single IRS between $S$ and $D$, its achievable rate varies with the distance $d$.   
	From Fig. \ref{fig:rate_location}, we observe that there exist two distance regions in which   the single IRS set-up outperforms the distributed IRSs  in terms of the achievable rate.  
	For instance, when the single IRS is placed nearer to $S$ or $D$, it achieves high achievable rates compared to that of the distributed IRS cases with fixed locations. 
	On the other hand, when the location of the single IRS is in between $d=40$\,m and $d=80$\,m (which is around the midpoint ($d=60$\,m)), the both distributed IRS set-ups with $\{N=2,L=64\}$ and $\{N=4,L=32\}$  outperform the single IRS case in terms of the achievable rate. The observations of  Fig. \ref{fig:rate_location} clearly reveal that an optimally positioned single IRS with $NL$ elements outperforms $N$   uniformly distributed IRSs  with  each having $L$ elements.
	Thus, the positioning of IRSs in a distributed set-up is a key factor in optimizing the performance gains.

\section{Conclusion}\label{sec:conclusion}
In this paper, the fundamental performance metrics of a distributed IRSs-aided communication system have been investigated. The optimal received SNR has been derived by controlling the phase-shifts of the distributed IRS elements. To facilitate a mathematically tractable performance analysis, this SNR has been statistically characterized  by deriving the tight approximations to the PDF and CDF for Nakagami-$m$ fading. 
Then, the outage probability, average SER, and achievable rate approximations/bounds   have been derived. Useful insights on the diversity order and array gain have been drawn by deriving the asymptotic outage probability and the average SER in  the high SNR regime. By virtue of an asymptotic rate analysis, it has been shown that the transmit power can be scaled inversely proportional to the square of the number of reflective elements, and in this operating regime, our lower and upper rate bounds have been shown to be asymptotically accurate. 
The detrimental effects of phase-shift quantization errors and  spatially-correlated fading at the IRSs have been investigated for the proposed distributed IRSs set-up.
A rigorous set of Monte-Carlo simulations has been presented to validate the accuracy of our analysis. The performance of the proposed distributed IRSs-aided set-up has been compared with   a conventional/direct  and a singe IRS-aided counterparts. Our analysis and numerical results reveal that   distributed IRSs deployment can be used to significantly enhance the performance of future wireless communications by recycling the existing EM waves via passive IRS reflective elements without generating new EM waves with active RF chains. Thus, the concept of  distributed IRSs-aided communications can be an energy and cost effective paradigm-shift in future wireless  designs.

\appendices
\section{The derivation of $\mu_{Y}$ in \eqref{eqn:mean}, $\sigma_{Y}^2$ in \eqref{eqn:var}, and PDF of $ Y$ in \eqref{eqn:pdf_Y}  }\label{app:Appendix0}
First, we define ${Y} = \sum_{n=1}^{N} \sum_{l=1}^{L} \eta_{nl} z_{nl}$, where $z_{nl} = \alpha_{h_{nl}}\alpha_{g_{nl}}$. Since $\alpha_{h_{nl}}$ and $\alpha_{g_{nl}}$ are independent Nakagami-$m$  RVs, the PDF of $z_{nl}$ can be evaluated as 
\vspace{-1mm}
\begin{eqnarray}\label{eqn:Apx_0_eqn_1}
	f_{z_{nl}}(x) &=& \int_{0}^{\infty} f_{\alpha_{h_{nl}}}(u) f_{\alpha_{g_{nl}}}(x/u) \times \frac{1}{|u|} du = \alpha x^{2m_g-1} \int_{0}^{\infty} u^{2(m_h-m_g)-1} \exp{-\frac{m_h u^2}{\xi_{h_n}} -\frac{m_g x^2}{\xi_{g_n}u^2}} du \nonumber \\
	&\stackrel{(a)}{=}& \frac{\alpha}{2} x^{2m_g-1} \int_{0}^{\infty} t^{m_h-m_g-1} \exp{-\frac{m_h t}{\xi_{h_n}} -\frac{m_g x^2}{\xi_{g_n} t}} dt 
	\stackrel{(b)}{=} \alpha' x^{m_h+m_g-1} \mathcal{K}_{m_h-m_g} \left(2x\sqrt{\frac{m_h m_g }{\xi_{h_n} \xi_{g_n}}}\right)\!, 
\end{eqnarray} 

\vspace{-1mm}
\noindent where $\alpha \!=\! 4 m_h^{m_{h}} m_g^{m_{g}} \!/\!\! \left(\Gamma\!(m_h) \Gamma\!(m_g) \xi_{h_n}^{m_h} \xi_{g_n}^{m_g}\!\right)$ and  $\alpha'\!\!=\! \alpha \!\left(\!m_g \xi_{h_n}\!/\! m_h \xi_{g_n}\!\right)^{\!\frac{m_h\!-\!m_g}{2}}$.
The step ($a$) is obtained  by letting $t \!=\! u^2$ and the step ($b$) is computed via  \cite[Eq. (3.471.9)]{Gradshteyn2007}. Then, the $n$th moment of $z_{nl}$ is derived as 
\vspace{-1mm}
\begin{eqnarray}\label{eqn:Apx_0_eqn_2}
	\E{z_{nl}^n} &=& \alpha' \int_{0}^{\infty} x^{m_h+m_g+n-1} \mathcal{K}_{m_h-m_g} \left(2x\sqrt{\frac{m_h m_g }{\xi_{h_n} \xi_{g_n}}}\right) dx \nonumber\\
	&\stackrel{(c)}{=}& \frac{\alpha'}{4}  \left(\frac{m_h m_g}{\xi_{h_n} \xi_{h_n}}\right)^{\frac{-m_h-m_g-n+1}{2}} \Gamma\left(\frac{2m_h+n}{2}\right) \Gamma\left(\frac{2m_g+n}{2}\right),  
\end{eqnarray} 

\vspace{-1mm}
\noindent where the step ($c$) is written via \cite[Eq. (6.561.16)]{Gradshteyn2007}.
The mean and  variance of $Y$ are given as
\vspace{-1mm}
\begin{eqnarray}
\!\!\!\!\!\!\!\!\!\!	\mu_{Y} &=& \E{Y} = \sum\limits_{n=1}^{N} \sum\limits_{l=1}^{L} \eta_{nl} \E{z_{nl}}, \label{eqn:Apx_0_eqn_3} \quad \text{and}\quad
	\sigma_{Y}^2  =  \Var{Y} = \sum\limits_{n=1}^{N} \sum\limits_{l=1}^{L} \eta_{nl}^2 \left(\E{z_{nl}^2} - \E{z_{nl}}^2\right).  \label{eqn:Apx_0_eqn_4}
\end{eqnarray} 

\vspace{-1mm}
\noindent By substituting \eqref{eqn:Apx_0_eqn_2} into \eqref{eqn:Apx_0_eqn_3}, $\mu_{Y}$ and $\sigma_{Y}^2$ can be computed as   \eqref{eqn:mean} and \eqref{eqn:var}, respectively. Then,  the PDF of $Y$ can be approximated by an one-sided Gaussian distribution as given in \eqref{eqn:pdf_Y} by invoking the central limit theorem  \cite{papoulis02}.

\section{The derivation of PDF of $\tilde R$ in \eqref{eqn:pdf_R}  }\label{app:Appendix1}
Since   $\alpha_u$ and $Y$ are independent RVs,   the PDF of $\tilde R = \alpha_u +\tilde Y$ can be derived as follows:
\vspace{-1mm}
\begin{eqnarray}\label{eqn:Apx_1_eqn_1}
	f_{\tilde R}(x) &=& \int_{0}^{\infty} f_u(u) f_{\tilde Y}(x-u) du = 
	2a^{m_u} \lambda \exp{-\frac{(x-\mu_{Y})^2}{2\sigma_{Y}^2}} \int_{0}^{\infty} u^{2m_u-1} \exp{-au^2+bu} du \nonumber \\
	&=& 2a^{m_u} \lambda \exp{-\frac{(x-\mu_{Y})^2}{2\sigma_{Y}^2}} \exp{\frac{b^2}{4a}}  \underbrace{\int_{0}^{\infty} u^{2m_u-1} \exp{-a \left(u-\frac{b}{2\sqrt{a}}\right)^2} du}_{I_{\tilde R}}, 
\end{eqnarray} 

\vspace{-1mm}
\noindent where $b \!=\! (x\!-\!\mu_{Y})\!/\!\sigma_{Y}^2$. Here,  $a$ and $\lambda$ are defined in \eqref{eqn:def_1}. 
The solution to $I_{\tilde R}$ in \eqref{eqn:Apx_1_eqn_1} is given by
\vspace{-1mm}
\begin{eqnarray}\label{eqn:Apx_1_eqn_2}
	I_{\tilde R} &=& \frac{1}{a^{m_u}} \int_{\frac{-b}{2\sqrt{a}}}^{\infty} \left(t+\frac{b}{2\sqrt{a}}\right)^{2m_u-1} \!\!\!\!\! \exp{-t^2} dt = \frac{1}{a^{m_u}} \sum_{k=0}^{2m_u-1} \binom{2m_u-1}{k} \left(\frac{b}{2\sqrt{a}}\right)^{2m_u-1-k} \int_{\frac{-b}{2\sqrt{a}}}^{\infty} t^{k} \exp{-t^2} dt \nonumber \\
	&\stackrel{(d)}{=}&  \frac{1}{2a^{m_u}}  \sum_{k=0}^{2m_u-1} \binom{2m_u-1}{k} \left(\frac{b}{2\sqrt{a}}\right)^{2m_u-1-k}  \Gamma\left(\frac{k+1}{2}, \frac{b^2}{4a}\right), 
\end{eqnarray} 

\vspace{-1mm}
\noindent where the step ($d$) is due to \cite[Eq. (2.33.10)]{Gradshteyn2007}. By substituting $(b)$    and \eqref{eqn:Apx_1_eqn_2} into  \eqref{eqn:Apx_1_eqn_1}, the PDF of $\tilde R$ is derived as  \eqref{eqn:pdf_R}.

\vspace{-3mm}
\section{The derivation of CDF of $\tilde R$ in \eqref{eqn:cdf_R}  }\label{app:Appendix2}
By substituting \eqref{eqn:pdf_R} into \eqref{eqn:cdf_R}, $I_{k}$ in \eqref{eqn:cdf_R} can be given as
\vspace{-1mm}
\begin{eqnarray}\label{eqn:Apx_2_eqn_1}
	I_{k} &=& \int_{x}^{\infty} \exp{-\Delta\left(\frac{u-\mu_Y}{q}\right)^2} \left(\frac{u-\mu_Y}{q}\right)^{2m_u-1-k} \Gamma \left(\frac{k+1}{2}, \left(\frac{u-\mu_Y}{q} \right)^2 \right) du \nonumber \\
	&\stackrel{(e)}{=} & q \int_{l_{min} }^{\infty} \exp{-\Delta t^2} t^{2m_u-1-k} \Gamma\left(\frac{k+1}{2},t^2\right) dt,
\end{eqnarray}

\vspace{-1mm}
\noindent where $q = 2\sigma_{Y}^2 \sqrt{a}$, $l_{min} = (x-\mu_{Y})/q$, and the step ($e$) is written by changing the variable of integration. Then,  we divide $I_{k}$ into two integrals: (i) $I_{k}^o$ for odd values of $k$ and (ii) $I_{k}^e$ for even values of $k$. Thus, for odd values of $k$ and when $x-\mu_{Y} < 0$, this integral can be written as
\vspace{-1mm}
\begin{eqnarray}\label{eqn:Apx_2_eqn_2}
	I_{k}^o &=& q \underbrace{\int_{l_{min}}^{0} \exp{-\Delta t^2} t^{2m_u-1-k} \Gamma\left(\frac{k+1}{2},t^2\right) dt}_{I_{k1}^o} + q \underbrace{\int_{0}^{\infty} \exp{-\Delta t^2} t^{2m_u-1-k} \Gamma\left(\frac{k+1}{2},t^2\right) dt}_{I_{k2}^o}.
\end{eqnarray} 
 Then, the first integral $I_{k1}^o$ in  \eqref{eqn:Apx_2_eqn_2} can be computed as 
\vspace{-1mm}
\begin{eqnarray}\label{eqn:Apx_2_eqn_3}
	I_{k1}^o &\stackrel{(f)}{=}& \int_{0}^{-l_{min}} \!\!\! \exp{-\Delta y^2} y^{2m_u-1-k} \Gamma\left(\frac{k+1}{2},y^2\right) dy  \stackrel{(g)}{=}  (\gamma_o-1)! \sum_{i=0}^{\gamma_o-1} \frac{1}{i!} \int_{0}^{-l_{min}}  \exp{-(\Delta+1) y^2} y^{2m_u-1-k} dy \nonumber \\
	&\stackrel{(h)}{=}& \frac{(\gamma_o-1 )!}{2} \sum\nolimits_{i=0}^{\gamma_o-1}  \frac{(\Delta+1)^{k/2-m_u-i}}{i!} \int_{0}^{l_{min}^2(\Delta+1)}  v^{m_u+i-k/2} \exp{-v}dv \nonumber \\
	&\stackrel{(i)}{=}& \frac{(\gamma_o-1 )!}{2} \sum\nolimits_{i=0}^{\gamma_o-1}  \frac{(\Delta+1)^{k/2-m_u-i}}{i!} \gamma\left(m_u+i-\frac{k}{2}, (\Delta+1)l_{min}^2 \right),   
\end{eqnarray}
 where  the step ($f$) and  step ($h$) are obtained by substituting $t\!=\!-y$ and $v \!=\! (\Delta+1)y^2$, respectively. The step ($g$) is computed via the fact that $\Gamma(n,x)=(n-1)! \exp{-x} \sum\nolimits_{m=0}^{n-1} x^m/m!$  \cite[Eq. (8.352.7)]{Gradshteyn2007}. The step ($i$) is written via  \cite[Eq. (8.350.1)]{Gradshteyn2007}. The second integral $I_{k2}^o$ in  \eqref{eqn:Apx_2_eqn_2} is computed as 
\vspace{-1mm}
\begin{eqnarray}\label{eqn:Apx_2_eqn_4}
	I_{k2}^o &=& \int_{0}^{\infty} \exp{-\Delta t^2} t^{2m_u-1-k} \Gamma\left(\frac{k+1}{2},t^2\right) dt \stackrel{(j)}{=} (\gamma_o-1)! \sum_{i=0}^{\gamma_o-1} \frac{1}{i!} \int_{0}^{\infty} t^{2m_u+2i-1-k} \exp{-(\Delta+1)t^2} dt \nonumber \\
	&\stackrel{(k)}{=}&  \frac{(\gamma_o-1)!}{2} \sum\nolimits_{i=0}^{\gamma_o-1} \frac{(\Delta+1)^{k/2-m_u-i}}{i!}  \Gamma\left(m_u+i-\frac{k}{2}\right),  
\end{eqnarray} 
 where the step ($j$) is derived  via a similar technique used in the step ($g$),  and the step ($k$) is evaluated by using \cite[Eq. (2.33.10)]{Gradshteyn2007}. Then, for $x-\mu_{Y} > 0$, $I_{k}^o$ can be calculated as
\begin{eqnarray}\label{eqn:Apx_2_eqn_5}
	I_{k}^o &=& \int_{l_{min}}^{\infty} \exp{-\Delta t^2} t^{2m_u-1-k} \Gamma\left(\frac{k+1}{2},t^2\right) dt =  (\gamma_o-1)! \sum_{i=0}^{\gamma_o-1} \frac{1}{i!} \int_{l_{min}}^{\infty} t^{2m_u+2i-1-k} \exp{-(\Delta+1)t^2} dt \nonumber\\
	&=&  \frac{(\gamma_o-1)!}{2} \sum\nolimits_{i=0}^{\gamma_o-1}  \frac{(\Delta+1)^{k/2-m_u-i}}{i!}  \Gamma\left(m_u+i-\frac{k}{2} ,(\Delta+1)l_{min}^2\right).  
\end{eqnarray} 
By combining \eqref{eqn:Apx_2_eqn_2} and \eqref{eqn:Apx_2_eqn_5},  $I_{k}^o$ can be derived  as  \eqref{eqn:I_mu_odd}. Then, for even values of $k$, the integral in \eqref{eqn:Apx_2_eqn_1} can be evaluated as 
\vspace{-1mm}
\begin{eqnarray}\label{eqn:Apx_2_eqn_6}
	I_{k}^e &=& q \int_{l_{min}}^{\infty} \exp{-\Delta t^2} t^{2m_u-1-k} \Gamma\left(\frac{k+1}{2},t^2\right) dt \\
	&\stackrel{(l)}{=}& \frac{(\gamma_e-1)!}{2} \sum_{j=0}^{\gamma_e-1}\left[ -\frac{\exp{-\Delta t^2} t^{2j}}{j! \Delta^{\gamma_e-1}} \Gamma\left(\frac{k+1}{2}, t^2\right) \right]_{l_{min}}^{\infty}  \!\!- (\gamma_e-1)! \sum_{j=0}^{\gamma_e-1} \frac{\Delta^{\gamma_e-1}}{j!} \int_{l_{min}}^{\infty}\!\! t^{k+2j} \exp{-(\Delta-1)t^2}dt \nonumber \\
	&=& \frac{(\gamma_e \!- \!1)!}{2} \sum_{j=0}^{\gamma_e-1}\! \frac{\exp{-\Delta l_{min}^2} l_{min}^{2j}}{j! \Delta^{\gamma_e-1}} \Gamma\!\left(\frac{k\!+\!1}{2}, l_{min}^2\right) \! -\! \frac{(\gamma_e\!-\!1)!}{2} \sum_{j=0}^{\gamma_e-1}\! \frac{\Delta^{\gamma_e-1}}{j!} \frac{\Gamma\left(j+\frac{k}{2}+\frac{1}{2}, (\Delta+1)l_{min}^2\right)}{(\Delta+1)^{j+k/2+1/2}}, \nonumber
\end{eqnarray} 
 where the step ($l$) is obtained  by using the  part-by-part integration technique.

\section{}\label{app:Appendix31}
\subsection{The Derivation of Expectation of $\tilde{\gamma}^*$ in \eqref{eqn:E_gamma_ub}  }\label{app:Appendix3_1}
The expectation term in  \eqref{eqn:E_gamma_ub}   can be computed as
\vspace{-1mm}
\begin{eqnarray}\label{eqn:Apx_3_eqn_1}
	\!\!\!\!\!\!\!\!\! \E{\tilde{\gamma}^*} \!&=&\! \E{\bar{\gamma} \tilde R^2} \!\!=\! \bar{\gamma}\E{\!(\alpha_u \!+\! \tilde Y)^2\!} \!\!=\! \bar{\gamma} \!  \sum_{n=0}^{2} \!\! \binom{2}{n} \E{\alpha_u^{(2-n)}} \!\E{\tilde Y^n} \!\!=\! \bar{\gamma} \!\left(\!\sigma_{u}^2 \!+\! \mu_{u}^2 \!+\! \sigma_{Y}^2 \!+\! \mu_{Y}^2 \!+\! 2 \mu_{u}\mu_{Y} \! \right)\!\!, 
\end{eqnarray}

\vspace{-1mm}
\noindent where $\mu_{Y}$, $\sigma_{Y}^2$, $\mu_{u}$, and $\sigma_{u}^2$ are given in \eqref{eqn:mean}, \eqref{eqn:var}, \eqref{eqn:u_mean}, and \eqref{eqn:u_var}, receptively.

\subsection{The Derivation of Variance of $\tilde{\gamma}^*$ in \eqref{eqn:Var_gamma_lb}  }\label{app:Appendix3_2}
The first expectation term in \eqref{eqn:Var_gamma_lb} can be evaluated as  
\vspace{-1mm}
\begin{eqnarray}\label{eqn:Apx_3_eqn_2}
	\E{\tilde R^4} &=& \E{(\alpha_u+ \tilde Y)^4} = \sum_{n=0}^{4} \binom{4}{n} \E{\alpha_u^{(4-n)}} \E{\tilde Y^n}. 
\end{eqnarray}

\vspace{-1mm}
\noindent Thus, the $n$th moment of $\alpha_u^n$ denoted by  $\E{\alpha_u^n}$ can be computed as
\vspace{-1mm}
\begin{eqnarray}\label{eqn:Apx_3_eqn_3}
	\!\!\!\!\! \E{\alpha_u^n} &=& \!\int_{0}^{\infty} \!\!\! x^n f_{u}(x) dx = \frac{2m_u^{m_u}}{\Gamma(m_u) \xi_u^{m_u}} \int_{0}^{\infty} \!\!\! x^{2m_u+n-1} \exp{-\frac{m_ux^2}{\xi_u}} dx \stackrel{(m)}{=} \frac{\xi_u^{n/2}}{\Gamma(m_u) m_u^{n/2}} \Gamma\left(m_u+n/2\right)\!,  
\end{eqnarray}

\vspace{-1mm}
\noindent where the step ($m$) is evaluated by using \cite[Eq. (2.33.10)]{Gradshteyn2007}. Then,   $\E{\tilde Y^n}$ can be derived as 
\vspace{-1mm}
\begin{eqnarray}\label{eqn:Apx_3_eqn_4}
	\E{\tilde Y^n} &=& \frac{\psi}{\sqrt{2\pi \sigma_{Y}^2}}  \int_{0}^{\infty} y^{n}  \exp{-\frac{(y-\mu_{Y})^2}{2\sigma_{Y}^2}} dy \stackrel{(n)}{=} \frac{\psi}{\sqrt{\pi }} \int_{\frac{-\mu_{Y}}{\sqrt{2\sigma_{Y}^2}}}^{\infty} \left(\sqrt{2\sigma_{Y}^2}t +\mu_{Y}\right)^n \exp{-t^2} dt \nonumber \\
	&\stackrel{(o)}{=}& \frac{\psi}{2\sqrt{\pi}} \sum_{i=0}^{n} \binom{n}{i}\left(\sqrt{2 \sigma_{Y}^2}\right)^{n-i} \mu_{Y}^i I\!\left(n-i, \frac{-\mu_{Y}}{2 \sigma_{Y}^2}\right), 
\end{eqnarray}
 where the step ($n$) is due to a changing of variable and  the step ($o$) is obtained by expanding $\left(\sqrt{2\sigma_{Y}^2}t +\mu_{Y}\right)^n$ based on $n$ value. Moreover, $I(\cdot,\cdot)$ is given in \eqref{eqn:I}.

\subsection{The Derivation of the Asymptotic Achievable Rate $(\mathcal{R}^{\infty})$ as $L\rightarrow \infty$ in \eqref{eqn:rate_asym}  }\label{app:Appendix3_3}
When the number of IRS elements grows without bound ($L \rightarrow \infty$), $\mu_{Y}$ in \eqref{eqn:mean} and $\sigma_{Y}^2$ in \eqref{eqn:var}  can be approximated as
\vspace{-1mm}
\begin{eqnarray}\label{eqn:Apx_3_3_eqn_1}
	\mu_{Y} &\approx& L \bar{\mu}_{Y} = L  \left(\sum_{n=1}^{N} \eta_{n} \sqrt{\frac{\xi_{h_n} \xi_{g_n}}{m_h m_g}} \frac{\Gamma\left(m_h + 1/2\right) \Gamma\left(m_g+1/2\right)}{\Gamma\left(m_h\right) \Gamma\left(m_g\right)} \right), \\
	\sigma_{Y}^2  &\approx& L \bar{\sigma}_{Y}^2 = L \left(\sum_{n=1}^{N}  \eta_{n}^2  \left({\frac{\xi_{h_n} \xi_{g_n}}{m_h m_g}} \right)  \frac{\Gamma\left(m_h+1\right) \Gamma\left(m_g+1\right)}{\Gamma\left(m_h\right) \Gamma\left(m_g\right)} -\bar{\mu}_Y^2\right),
\end{eqnarray}
  where  $\eta_{n} = \eta_{nl}$. Thus, the SNR term in $\mathcal{R}_{ub}$ \eqref{eqn:rate_ub_sub} can be asymptotically evaluated as
\vspace{-1mm}
\begin{eqnarray}\label{eqn:Apx_3_3_eqn_2}
	\lim_{L \rightarrow \infty} \gamma_{ub} &=& \lim_{L \rightarrow \infty} \bar{\gamma} L^2 \left(\frac{\sigma_{u}^2}{L^2} +\frac{\bar{\sigma}_{Y}^2}{L} + \frac{2\mu_{u} \bar{\mu}_{Y}}{L} + \frac{\mu_{u}^2}{L^2} + \bar{\mu}_{Y}^2 \right) = \bar{\gamma}_{E} \bar{\mu}_{Y}^2,
\end{eqnarray}
where $\bar{\gamma}_{E} = P_{E}/\sigma^2_w$ and $\lim_{L\rightarrow \infty} P = P_E/L^2$. Similarly, in the asymptotic regime, the SNR   term in $\mathcal{R}_{lb}$ \eqref{eqn:rate_lb_sub} can be derived as
\vspace{-1mm}
\begin{eqnarray}\label{eqn:Apx_3_3_eqn_3}
	\!\!\!\!\!\!\!\!\!\!	\lim_{L \rightarrow \infty} \!\gamma_{lb} &=&\! \lim_{L \rightarrow \infty} \bar{\gamma}\!\! \left(\!\! \frac{ L^6 \left(\frac{\sigma_{u}^2}{L^2} +\frac{\bar{\sigma}_{Y}^2}{L} + \frac{2\mu_{u} \bar{\mu}_{Y}}{L} + \frac{\mu_{u}^2}{L^2} + \bar{\mu}_{Y}^2 \right)^{\!\!3}}{ L^4 \left(\frac{\psi}{2\sqrt{\pi}}  \bar{\mu}_{Y}^4   I\left(0, \frac{-\sqrt{L}\bar{\mu}_{Y}}{ \sqrt{2 \bar{\sigma}_{Y}^2}}\right) + \frac{\Upsilon(L^3)}{L^4} \right) } \!\!\right) 
	\!\!=\!\! \lim_{L \rightarrow \infty} \bar{\gamma} L^2\!\! \left( \!\frac{ \bar{\mu}_{Y}^2}{ \frac{\psi}{2\sqrt{\pi}}     I\left(0, -\infty \right)   } \!\!\right)  
	\!\! \stackrel{(p)}{=} \!\! \bar{\gamma}_{E} \bar{\mu}_{Y}^2,
\end{eqnarray}
  where $\Upsilon(L^3)$ is lower order ($L<4$) terms of the expansion of the SNR in   \eqref{eqn:rate_lb_sub}. Moreover, the step ($p$) is written by using the fact that  $\lim_{L \rightarrow \infty} \psi = \lim_{L \rightarrow \infty} 1/\mathcal{Q}\left(-\sqrt{L}\bar{\mu}_Y/\bar{\sigma}_{Y}\right) = \lim_{L \rightarrow \infty} 1/\mathcal{Q}\left(-\infty\right) =1$ and $I\left(0, -\infty \right)  = 2 \Gamma(1/2) = 2 \sqrt{\pi}$. Then, the asymptotic achievable rate $\mathcal{R}^{\infty}$ can be derived as given in \eqref{eqn:rate_asym}.

\section{The derivation of the average SER in \eqref{eqn:avg_ber}  }\label{app:Appendix3}

The inner expectation term with respect to $\tilde Y$ in \eqref{eqn:ber_alt} can be evaluated via a tight approximation \cite{Andrew2013} for  the Gaussian $Q$-function: $\mathcal{Q}(x) \!\approx\!  \Exp{-cx^2-dx}/2$, where $c\!=\!0.374$ and $d \!=\! 0.777$   as 
\begin{eqnarray}\label{eqn:Apx_32_eqn_1}
	\!\!\!\!\!\!\!\!\! \E[\tilde{Y}]{\mathcal{Q}\left( \vartheta_b (\alpha_u+\tilde{Y})\right)} &=& \frac{1}{2} \int_{0}^{\infty} \exp{-cx^2-dx} f_{\tilde{Y}}(y) dy = \frac{\psi \exp{-c\vartheta_b^2\alpha_u^2 -d\vartheta_b\alpha_u - \frac{\mu_{Y}^2}{2 \sigma_{Y}^2}}}{2 \sqrt{2 \pi \sigma_{Y}^2}} \underbrace{\int_{0}^{\infty} \exp{-a_1 y^2 - a_2 y} dy}_{I_{\tilde{Y}}},
\end{eqnarray}
 where $a_1 = c\vartheta_b^2 + 1/2\sigma_{Y}^2$ and $a_2 = 2c \vartheta_b^2 \alpha_u + d \vartheta_b - \mu_{Y}/\sigma_{Y}^2$. Then, the integral $I_{\tilde{Y}}$ in \eqref{eqn:Apx_32_eqn_1} can be evaluated by first using a substitution $t=y+ a_2/(2a_1)$ and then invoking \cite[Eq. (2.33.16)]{Gradshteyn2007}  
\vspace{-0mm}
\begin{eqnarray}\label{eqn:Apx_4_eqn_2}
	I_{\tilde{Y}} &=& \Exp{\frac{a_2^2}{4a_1^2}} \int_{0}^{\infty} \Exp{-a_1 \left(y+ \frac{a_2}{2a_1}\right)^2} dy = \Exp{\frac{a_2^2}{4a_1^2}} \int_{a_2/2a_1}^{\infty} \Exp{-a_1 t^2} dt \nonumber \\
	&=& \frac{1}{2} \sqrt{\frac{\pi}{a_1}} \Exp{\frac{a_2^2}{4a_1^2}} \errc{\frac{a_2}{2 \sqrt{a_1}}}.
\end{eqnarray}
  Then, by substituting $a_2$ and \eqref{eqn:Apx_4_eqn_2} into \eqref{eqn:Apx_32_eqn_1}, the inner expectation in \eqref{eqn:ber_alt} can be rewritten as
\vspace{-0mm}
\begin{eqnarray}\label{eqn:Apx_4_eqn_3}
	\!\!\!\!\!\!\!\! \E[\tilde{Y}]{\mathcal{Q}\left( \vartheta_b (\alpha_u\!+\!\tilde{Y})\right)} \!&=&\!  \frac{\psi \Exp{\!-c\vartheta_b^2\alpha_u^2 -d\vartheta_b\alpha_u - \frac{\mu_{Y}^2}{2 \sigma_{Y}^2}}}{4 \sqrt{2 \sigma_{Y}^2 a_1}}  \Exp{-v' \alpha_u^2 -u_1 \alpha_u} \errc{s \alpha_u \!+ r}\!,
\end{eqnarray}
  where $v' = v_1 - m_u/P_u$. Moreover,  $s$, $r$, $u_1$, and $v_1$ are defined in \eqref{eqn:definitions}. 
 By substituting \eqref{eqn:Apx_4_eqn_3} into \eqref{eqn:ber_alt}, the outer expectation with  respect to $\alpha_u$ can be written  as
 \vspace{-0mm} 
\begin{eqnarray}\label{eqn:Apx_4_eqn_4}
	\!\!\!\!\!\!\!\!\!\! \E[\alpha_u]{\E[\tilde{Y}]{\mathcal{Q}\left(  \vartheta_b ( \alpha_u+\tilde{Y} ) \right)}} &=& \int_{0}^{\infty}  \E[\tilde{Y}]{\mathcal{Q}\left( \vartheta_b (\alpha_u+\tilde{Y})\right)} f_u(\alpha_u) d\alpha_u \nonumber \\
	&=&  2\sqrt{2\pi} B  \underbrace{\int_{0}^{\infty}  \alpha_u^{2m_u -1} \exp{-v_1\left(\alpha_u + \frac{u_1}{2v_1}\right)^2} \mathcal{Q} \left( \sqrt{2}(s\alpha_u+r)\right)  d\alpha_u }_{I_{\alpha_u}},
\end{eqnarray}
  where $B$ is defined in \eqref{eqn:definitions}. 
Then, the integral $I_{\alpha_u}$ in \eqref{eqn:Apx_4_eqn_4} can be evaluated as
\vspace{-1mm}
\begin{eqnarray}\label{eqn:Apx_4_eqn_5}
	&&\!\!\!\!\!\!\!\!\!\!\!I_{\alpha_u} = \int_{0}^{\infty} \alpha_u^{2m_u -1} \exp{-v_1\left(\alpha_u + \frac{u_1}{2v_1}\right)^2} 	\mathcal{Q}\left(\sqrt{2}(s\alpha_u+r)\right) d\alpha_u  \\
	&\stackrel{(q)}{=}& \frac{-1}{2} \sum_{k=0}^{2m_u-1} \!\!\left(\frac{-u_1}{2v_1} \right)^{\!\!2m_u-1-k}  \Gamma(\gamma_b)\!   \left[ \mathcal{Q}\left( \sqrt{2}(s\alpha_u+r) \right) 
	\exp{-v_1\left(\alpha_u + \frac{u_1}{2v_1}\right)^2} \sum_{i=0}^{\gamma_b-1}  \frac{\left(\alpha_u+ u_1/2v_1\right)^{2i}}{i! v_1^{\gamma_b-i}} \right]_0^{\infty} \nonumber \\
	&&
	- \frac{s}{2\sqrt{\pi}} \!\!\sum_{k=0}^{2m_u-1} \!\!\left(\frac{-u_1}{2v_1}\right)^{\!2m_u-1-k}\!\! \Gamma(\gamma_b) \sum_{i=0}^{\gamma_b-1} \frac{1}{i! v_1^{\gamma_b-i} }  \int_{0}^{\infty}\!\! \left(\alpha_u+ \frac{u_1}{2v_1}\right)^{\!\!2i} \!\!\exp{-v_1\left(\alpha_u+\frac{u_1}{2v_1}\right)^2 - (s\alpha_u+r)^2} d\alpha_u \nonumber \\
	&=& \frac{\mathcal{Q}\left( \sqrt{2}r \right)}{2} \sum_{k=0}^{2m_u-1} \left(\frac{-u_1}{2v_1} \right)^{2m_u-1-k}  \Gamma(\gamma_b)  \sum_{i=0}^{\gamma_b-1} \frac{\left( u_1/2v_1\right)^{2i}}{i! v_1^{\gamma_b-i}} \nonumber \\
	&&
	- \frac{s}{2\sqrt{\pi}} \sum_{k=0}^{2m_u-1} \left(\frac{-u_1}{2v_1}\right)^{\!\!2m_u-1-k} \!\!\Gamma(\gamma_b)\! \sum_{i=0}^{\gamma_b-1} \frac{1}{i! v_1^{\gamma_b-i} }  \underbrace{\int_{0}^{\infty} \!\!\left(\!\alpha_u \!+\! \frac{u_1}{2v_1}\!\right)^{\!2i} \exp{-v_1\left(\alpha_u+\frac{u_1}{2v_1}\right)^2 - (s\alpha_u+r)^2} d\alpha_u }_{I_{\alpha_u'}}, \nonumber
\end{eqnarray}

\vspace{-1mm}
\noindent where $\gamma_b = (k+1)/2$. The  step ($q$) is written by using the part-by-part integration technique. Then, the integral $I_{\alpha_u'}$ can be rearranged as follows:
\vspace{-1mm}
\begin{eqnarray}\label{eqn:Apx_4_eqn_6}
	I_{\alpha_u'} &=& \exp{\frac{u_2^2}{4v_2} - \frac{u_1^2}{4v_1}-r} \int_{0}^{\infty} \left(\alpha_u+\frac{u_1}{2v_1}\right)^{2i} \exp{-v_2\left(\alpha_u+\frac{u_2}{2v_2}\right)^2} d\alpha_u \nonumber \\
	&\stackrel{(r)}{=}& \exp{\frac{u_2^2}{4v_2} - \frac{u_1^2}{4v_1}-r} \sum_{j=0}^{2i} q_b^{2i-j} \int_{\frac{u_2}{2v_2}}^{\infty} t^j \exp{-v_2t^2} dt = \exp{\frac{u_2^2}{4v_2} - \frac{u_1^2}{4v_1}-r} \sum_{j=0}^{2i} q_b^{2i-j} I_b,
\end{eqnarray}

\vspace{-1mm}
\noindent where $q_b = u_1/2v_1 - u_2/2v_2$  and $I_b$ is defined in (\ref{eqn:I_b}). Further, the step ($r$) is obtained via the substitution $t= \alpha_u + u_2/(2v_2)$. Finally, $u_2$ and  $v_2$  are defined in \eqref{eqn:definitions}. 

\section{The derivation of ${P}^{\infty}_{out}$ in \eqref{eqn:asym_outage} and $\bar{P}_{e}^{\infty}$ in \eqref{eqn:asym_ber}  }\label{app:Appendix5}

From (\ref{eqn:Apx_0_eqn_1}) in Appendix \ref{app:Appendix0}, the PDF of $z_{nl} = \alpha_{h_{nl}}\alpha_{g_{nl}}$ can be written as 
\vspace{-1mm}
\begin{eqnarray}\label{eqn:Apx_5_eqn_1}
	f_{z_{nl}}(x) &=&  \alpha' x^{m_s+m_l-1} \mathcal{K}_{m_s-m_l} \left(2x\sqrt{{m_s m_l }/{(\xi_{s_n} \xi_{l_n})}}\right), 
\end{eqnarray} 

\vspace{-1mm}
\noindent where $\alpha'$ is given in \eqref{eqn:alpha_dash}, $m_s = \min{(m_h,m_g)}$, and $m_l= \max{(m_h,m_g)}$. Moreover, $\xi_{s_n}$ and $\xi_{l_n}$ are the scaling parameters of the respective channel models \eqref{eqn:channels}. 
The moment generating function (MGF) of $z_{nl}$ can be derived by evaluating the Laplace transform of $f_{z_{nl}}(x)$ as  \cite{papoulis02}
\vspace{-0mm}
\begin{eqnarray}\label{eqn:Apx_5_eqn_2}
	\mathcal{M}_{z_{nl}}(s) &=& \!   \int_{0}^{\infty}\!\!\! f_{z_{nl}} (x)\mathrm{exp}(-sx) dx=   \alpha' \!\!\int_{0}^{\infty} \!\!\!x^{m_s+m_l-1} \exp{-sx} \mathcal{K}_{m_s-m_l} \left(2x\sqrt{{m_s m_l }/{(\xi_{s_n} \xi_{l_n})}}\right) dx  \\
	&&\!\!\!\!\!\!\!\!\!\!\!\!\!\!\!\!\!\!\!
	\stackrel{(s)}{=} \! \frac{\alpha' \sqrt{\pi} \left(2b_n\right)^{(m_s-m_l)} \Gamma(2m_s) \Gamma(2m_l)}{\Gamma(m_s+m_l+1/2)} \left(s \!+\! b_n\right)^{2m_s} F\left(\!2m_s, m_s\!-\!m_l\!+\!\frac{1}{2};m_s\!+\!m_2\!+\!\frac{1}{2};\frac{s\!-\!b_n}{s\!+\!b_{n}}\right), \nonumber 
\end{eqnarray} 
where $b_n = 2 \sqrt{m_s m_l/ (\xi_{h_n} \xi_{g_n})}$.
In \eqref{eqn:Apx_5_eqn_2}, the step $(s)$ is derived by using \cite[Eq. (6.621.3)]{Gradshteyn2007}.
The behavior of the PDF of $z_{nl}$ at the origin is governed by the asymptotic value of  $\mathcal{M}_{z_{nl}}(s)$ as $s\rightarrow\infty $ \cite{Wang2003a}. To this end, 
by using the fact that $\lim_{s \rightarrow \infty}$ $({s-b_n})/({s+b_n}) \rightarrow 1$, and we invoke \cite[Eq. (15.4.21)]{Lozier2003} for $m_s=m_l$ and \cite[Eq. (9.122.1)]{Gradshteyn2007} for $m_s<m_l$ to evaluate $\mathcal{M}_{z_{nl}}(s)$ when $s \rightarrow \infty$ as follows:
\vspace{-7mm}
\begin{eqnarray}\label{eqn:Apx_5_eqn_3}
	\mathcal{M}_{z_{nl}}^{\infty}(s) &=&  \theta_{n} s^{-2m_s},
\end{eqnarray} 
where $\theta_{n}$ is defined in \eqref{eqn:theta_n}. Thus, the MGF of ${Y} = \sum_{n=1}^{N} \sum_{l=1}^{L} \eta_{nl} z_{nl}$ as  $s \rightarrow \infty$ is given by
\vspace{-1mm}
\begin{eqnarray}\label{eqn:Apx_5_eqn_4}
	\mathcal{M}_{Y}^{\infty}(s) &=& \prod\nolimits_{n=1}^{N} \prod\nolimits_{l=1}^{L} \eta_{nl} \theta_{n} s^{-2m_s} = \Phi(N,L) s^{-2m_sNL},
\end{eqnarray} 
  where $\Phi(N,L)$ is given in \eqref{eqn:phi}. The PDF of the direct channel envelope $(u)$ is expanded at the origin via the Maclaurin series  expansion of the exponential function  \cite[Eq. (0.318.2)]{Gradshteyn2007} as
\vspace{-1mm}
\begin{eqnarray}\label{eqn:Apx_5_eqn_5}
	f_{\alpha_u}(x) &=& \frac{2m_u^{m_u} x^{2m_u-1}}{\Gamma(m_u)\xi_u^{m_u}} \Exp{\frac{- x^2}{\xi_u}} = \frac{2m_u^{m_u} x^{2m_u-1}}{\Gamma(m_u)\xi_u^{m_u}} \sum_{k=0}^{\infty} \frac{1}{k!} \left(\frac{-x^2}{\xi_u}\right)^k \nonumber \\
	&=& \frac{2m_u^{m_u} x^{2m_u-1}}{\Gamma(m_u)\xi_u^{m_u}} \left(1 - \frac{x^2}{\xi_u} + \frac{x^4}{2\xi_u^2} - \frac{x^6}{6\xi_u^3} + \cdots\right).
\end{eqnarray} 
 Then, the PDF of the direct channel  envelope $(u)$ for $x \rightarrow 0^+$ can be approximated as
\vspace{-1mm}
\begin{eqnarray}\label{eqn:Apx_5_eqn_6}
	f_{\alpha_u}^{0^+}(x) &=& \frac{2m_u^{m_u} x^{2m_u-1}}{\Gamma(m_u)\xi_u^{m_u}} +  \mathcal{O} \left(x^{2m_u}\right).
\end{eqnarray} 
  Via the Laplace transform  of \eqref{eqn:Apx_5_eqn_6},  we obtain the   asymptotic MGF of $\alpha_u$  as \cite[Eq. (17.13.2)]{Gradshteyn2007} 
\vspace{-2mm}
\begin{eqnarray}\label{eqn:Apx_5_eqn_7}
	\mathcal{M}_{\alpha_u}^{\infty}(s) &=& \frac{2m_u^{m_u} \Gamma(2m_u) }{\Gamma(m_u)\xi_u^{m_u}} s^{-2m_u}.
\end{eqnarray}
  Then, the asymptotic MGF of $R=\alpha_u + Y$ for $s \rightarrow \infty$ can be derived as
\vspace{-0mm}
\begin{eqnarray}\label{eqn:Apx_5_eqn_8}
	\mathcal{M}_{R}^{\infty}(s) &=& \mathcal{M}_{Y}^{\infty}(s) \mathcal{M}_{\alpha_u}^{\infty}(s) = \frac{\Phi(N,L) 2m_u^{m_u} \Gamma(2m_u) }{\Gamma(m_u)\xi_u^{m_u}} s^{-2m_sNL-2m_u},
\end{eqnarray}
 where $\Phi(N,L)$ is defined in \eqref{eqn:phi}.
By taking inverse Laplace transform of \eqref{eqn:Apx_5_eqn_8} \cite{Gradshteyn2007}, the PDF of $R$ for $x \rightarrow 0^+$ can be computed as
\vspace{-0mm}
\begin{eqnarray}\label{eqn:Apx_5_eqn_9}
	f_R^{0^+}(x) &\stackrel{(t)}{=}&  \frac{\Phi(N,L) 2m_u^{m_u} \Gamma(2m_u) }{\Gamma(m_u)\xi_u^{m_u} \Gamma(2m_sNL+2m_u)} x^{2m_sNL+2m_u-1} +  \mathcal{O} \left(x^{2m_sNL+2m_u}\right),
\end{eqnarray}
 where the step $(t)$ is written via \cite[Eq. (17.13.3)]{Gradshteyn2007}. Next, the CDF of $R$ for $x \!\rightarrow\! 0^+$ is given as 
\begin{eqnarray}\label{eqn:Apx_5_eqn_10}
	\!\!F_R^{0^+}(x) &=& \int_{0}^{x}\!\!\! f_R^{0+}(u) du \!=\!  \int_{0}^{x} \!\! \frac{\Phi(N,L) 2m_u^{m_u} \Gamma(2m_u) }{\Gamma(m_u)\xi_u^{m_u} \Gamma(2m_sNL+2m_u)} u^{2m_sNL+2m_u-1}  du +  \mathcal{O} \left(u^{2m_sNL+2m_u}\right) \nonumber \\
	&=& \frac{\Phi(N,L) m_u^{m_u} \Gamma(2m_u) }{G_d \Gamma(m_u)\xi_u^{m_u} \Gamma(2G_d) } x^{2G_d} + \mathcal{O} \left(x^{2G_d+1}\right),
\end{eqnarray}
  where $G_d = m_sNL+m_u$. Then, the CDF of $\gamma^* = \bar \gamma R^2$ for $x \rightarrow 0^+$ is approximated as \cite{papoulis02}
\vspace{-1mm}
\begin{eqnarray}\label{eqn:Apx_5_eqn_11}
\!\!\!\!\!	\!\!\!\!\!\!\! F_{\gamma^*}^{0^+}(x) &=&  \mathrm{Pr}\left(\gamma^*\leq x\right) \approx F_{\tilde R}^{0^+}\left(\sqrt{x/\bar{\gamma}} \right) = \frac{\Phi(N,L) m_u^{m_u} \Gamma(2m_u) }{G_d \Gamma(m_u)\xi_u^{m_u} \Gamma(2G_d) } \left(\frac{x}{\bar{\gamma}}\right)^{G_d} \!\!\!+ \mathcal{O}\! \left(\!\left(\frac{x}{\bar{\gamma}}\right)^{\!\!G_d+1}\right)\!.
\end{eqnarray}
  Thereby, the asymptotic outage probability can be computed as
\vspace{-0mm} 
\begin{eqnarray}\label{eqn:Apx_5_eqn_11_1}
	{P}^{\infty}_{out} \approx F_{\gamma^*}^{0^+}(\gamma_{th}) = \Omega \Phi(N,L)   \left(\frac{\gamma_{th}}{\bar{\gamma}}\right)^{G_d} + \mathcal{O} \left( \left(\frac{\gamma_{th}}{\bar{\gamma}}\right)^{G_d+1}\right),
\end{eqnarray}

\vspace{-1mm}
\noindent where $\Omega$ is given in \eqref{eqn:omega}.
Finally, the asymptotic average BER can be derived as follows:
\vspace{-1mm}
\begin{eqnarray}\label{eqn:Apx_5_eqn_12}
	P_e^{\infty} &=&  \int_{0}^{\infty} \omega \mathcal{Q}\left(\sqrt{\vartheta x}\right) f_{\gamma^*}^{0^+}(x) dx = \frac{\omega}{2} \sqrt{\frac{\vartheta}{2 \pi}}  \int_{0}^{\infty} x^{-1/2} \Exp{\frac{-\vartheta x}{2}} F_{\gamma^*}^{0^+}(x) dx \nonumber \\
	&=& \frac{\omega}{2} \sqrt{\frac{\vartheta}{2 \pi}} \frac{\Phi(N,L) m_u^{m_u} \Gamma(2m_u) }{G_d \Gamma(m_u)\xi_u^{m_u} \Gamma(2G_d) } \bar{\gamma}^{-G_d} \int_{0}^{\infty} x^{G_d-1/2} \Exp{\frac{-\vartheta x}{2}} dx + \mathcal{O}\! \left(\!\left(\frac{x}{\bar{\gamma}}\right)^{G_d+1}\right) \nonumber \\
	&\stackrel{(u)}{=}& \Lambda \Phi(N,L)  \bar{\gamma}^{-G_d} + \mathcal{O} \left(\bar{\gamma}^{-(G_d+1)}\right),
\end{eqnarray} 

\vspace{-1mm}
\noindent where $\Lambda$ is given in \eqref{eqn:lambda} and the step $(u)$ is computed via \cite[Eq. (3.326.2)]{Gradshteyn2007}.

\vspace{-5mm}
  \linespread{1.6}
\bibliographystyle{IEEEtran}
\bibliography{IEEEabrv,References_1}

\end{document}